\RequirePackage{fix-cm}
\documentclass[twocolumn,epjc3]{svjour3}

\smartqed  % flush right qed marks, e.g. at end of proof
\RequirePackage{graphicx}
\usepackage{mathptmx, courier, pifont}
\usepackage[scaled=0.92]{helvet}
\usepackage[T1]{fontenc}
\usepackage{textcomp}

\usepackage[pdftex,colorlinks=true]{hyperref}
\def\bra{\,<\!} \def\ket{\!>\,} \def\ack{\,|\,}

\journalname{Eur. Phys. J. A}
\begin{document}
\sloppy

\title{High-Spin Doublet Band Structures in odd-odd $^{194-200}$Tl isotopes
}


\author{S.~Jehangir\thanksref{e1,addr4}
         \and
         I. Maqbool\thanksref{addr2}
         \and
         G.H. Bhat\thanksref{e2,addr1,addr3}
        \and
        J.A. Sheikh\thanksref{e3,addr2,addr3}
        \and
        R. Palit\thanksref{addr4}
        \and
        N. Rather\thanksref{addr5}
}

\thankstext{e1}{e-mail: sheikhahmad.phy@gmail.com}
\thankstext{e2}{e-mail: gwhr.bhat@gmail.com}
\thankstext{e3}{e-mail: sjaphysics@gmail.com}


\institute{Department of Physics, SP College  Srinagar, Jammu and Kashmir, 190 001, India\label{addr1}
  \and
  Department of Physics, University of Kashmir, Srinagar,190 006, India \label{addr2}
  \and
Cluster University Srinagar, Jammu and Kashmir, Srinagar, Goji
Bagh, 190 008, India \label{addr3}
\and
Department of Nuclear and Atomic Physics, Tata Institute of Fundamental Research, Mumbai - 400 005, India\label{addr4}
\and
Department of Physics, Islamic University of Science and Technology,  Jammu and Kashmir, 192 122, India\label{addr5}
}

\date{Received: date / Accepted: date}

\maketitle

\begin{abstract}
The basis space in the triaxial projected shell model (TPSM) approach is generalized for
odd-odd nuclei to include two-neutron and two-proton configurations on
the basic one-neutron coupled to one-proton quasiparticle state. The
generalization allows to investigate odd-odd nuclei beyond the band
crossing region and as a first application of this development, high-spin band
structures recently observed in odd-odd $^{194-200}$Tl isotopes are
investigated. In some of these isotopes, the  doublet band
structures observed after the band crossing have been conjuctured to arise from the spontaneous
breaking of the chiral symmetry.  The driving configuration of the
chiral symmetry in these odd-odd isotopes is one-proton and
three-neutrons rather than the basic one-proton and one-neutron as
already observed in many other nuclei. It is demonstrated using the TPSM approach
that energy differences of the doublet bands in $^{194}$Tl and
$^{198}$Tl  are, indeed,  small. However,
the differences in the calculated transition probabilities
are somewhat larger than what is expected in the
chiral symmetry limit. Experimental data on the transition
probabilities is needed to shed light on the chiral nature of the doublet bands.
\end{abstract}


%\begin{keyword}

%Isotopic identification of fission fragments, neutron rich Rh
%isotopes, triaxial projected shell model, nuclear triaxiality

\PACS  21.60.Cs, 21.10.Hw, 21.10.Ky, 27.50.+e

%\end{keyword}
%\end{frontmatter}

\maketitle

\section{Introduction}
High-spin spectroscopy has played a pivotal role to unravel the 
structure of atomic nuclei at high angular momentum and 
excitation energy \cite{BM75}. The advancements in the spectroscopic techniques
have allowed to probe the properties of nuclei in regions  that were
hitherto inaccessible. The band structures have been identified up to
very high angular momentm and excitation energy and data has revealed 
interesting phase and shape transitions \cite{bm8}.  In odd-odd nuclei, the
energy spectrum is quite rich as compared to even-even and odd-mass
nuclei due to extra neutron-proton coupling. Recently, in odd-odd
$^{194-200}$Tl isotopes, band structures have been populated beyond the
first band crossing and it has been spectulated that near degeneracy
of the yrast and the side band energies, observed in some of these
isotopes, may be a consequence of the chiral symmetry breaking
mechanism \cite{sb17,pl13,HP12,ea08}. In these isotopes,
band structures have one-proton and
three-neutron, $\pi h_{9/2} \otimes \nu i^{-3}_{13/2} $  configuration
at higher spins and one-proton and one-neutron,  $\pi h_{9/2} \otimes
\nu i^{-1}_{13/2} $ configuration at lower spins. 

Chiral symmetry has been studied quite extensively in triaxial deformed
nuclei \cite{SF01,TS04,CHI,TS06}. In the original work, the chiral symmetry mechanism was
proposed for odd-odd nuclei with the angular-momentum of the
odd-proton and odd-neutron  aligned towards short- and long-axis,
and the angular-momentum of the deformed core projected towards the intermediate
axis \cite{SF97}. As the clock-wise and the counter clock-wise rotation of the 
three orthogonal vectors are equivalent, this results into doublet band
structures with identical spectroscopic properties. Doublet bands with
similar properties have been observed in several mass regions of the
periodic table \cite{CV04,TS03,KS01,AA01,SU03}, 
  including the A$\sim$ 80 region in recent works\cite{SUU1,SUU3,SUU}. Chiral symmetry 
breaking has also been proposed in odd-mass \cite{SU03,SU02} and 
even-even nuclei \cite{SZ05}. 
Theoretically, several nuclear models have been employed to
investigate the chiral symmetry breaking mechanism, which 
include both microscopic and phenomenological models
\cite{VD00,PO04,PRM,RPA1,RPA2}.
Triaxial
projected shell model (TPSM) approach, a semi-microscopic
model, has been used to elucidate the band structures in deformed
and transitional nuclei  \cite{JS99}. It has been demonstrated
 that the properties of
the observed doublet bands are reproduced quite successfully using the
TPSM approach \cite{JG12}. The advantage of the TPSM
approach is that computational needs are quite modest and it is
possible to perform systematic studies in a reasonable time frame. As a
matter of fact several systematic studies have been performed using
the TPSM approach \cite{bh14a,bh15b,Chanli15,Chanli16,Js16}. 

For odd-odd nuclei, the basis space in the TPSM approach is composed of 
one-neutron coupled to one-proton quasiparticle configurations \cite{JG12}. This
basis space is obviously quite restrictive and allows to study only
low-lying states in odd-odd nuclei. To study high-spin states
in odd-odd nuclei, around and beyond the band crossing, it is
imperative to include two-neutron and two-proton states coupled to
the basic one-neutron plus one-proton states. This extension of the
model space will allow to probe the high-spin band structures 
beyond the first band crossing. 
In the present work,
the model space has been expanded
to include two-neutron and two-proton quasiparticle configuration
over the primary one-proton plus one-neutron configuration. As a first
application of this development, the properties of the observed
band structures in odd-odd $^{194-200}$Tl nuclei are investigated.
In these isotopes, doublet band structures, observed above first
band crossing having four-quasiparticle configuration, are predicted
to originate from the chiral symmetry breaking mechanism. These shall
be first examples of the manifestation of chiral symmetry in odd-odd 
with four-quasiparticle structures. In all earlier studies of odd-odd nuclei, chiral
bands have been identified with one-proton and one-neutron
configuration.
The present work
is organised in the following manner. In the next section, the TPSM
approach is presented with the expanded model space. In section III,
the results are presented and discussed. Finally, the present work is 
summarized and concluded in section IV.

\section{Triaxial Projected Shell Model Approach}
In recent years, the triaxial projected shell model
approach has been shown to reproduce the properties
of deformed odd-odd nuclei quite well. In the earlier version, the basis space for
odd-odd nuclei was composed of one-neutron coupled to one-proton only \cite{JG12}. In
the present work, the basis space is generalized to include 
two-neutron and two-proton quasiparticle configurations coupled to the
basic one-neutron $\otimes$ one-proton configuration. The generalized
TPSM basis space  is given by $:$
\begin{equation}
\begin{array}{r}
%\hat P^I_{MK}\ack\Phi\ket;\\
~~\hat P^I_{MK}~a^\dagger_{\nu_1} a^\dagger_{\pi_1} \ack\Phi\ket;\\
~~\hat P^I_{MK}~a^\dagger_{\nu_1}a^\dagger_{\nu_2} a^\dagger_{\nu_3}a^\dagger_{\pi_1} \ack\Phi\ket;\\
~~\hat P^I_{MK}~a^\dagger_{\nu_1} a^\dagger_{\pi_1}a^\dagger_{\pi_2} a^\dagger_{\pi_3} \ack\Phi\ket,
\label{intrinsic}
\end{array}
\end{equation}
where $P^I_{MK}$ is the three-dimensional
angular-momentum-projection operator \cite{RS80}.
The triaxial quasi-particle (qp) vacuum configuration, $ | \Phi \ket $, in Eq.~(\ref{intrinsic}) is
constructed through diagonalization of the deformed Nilsson 
Hamiltonian and a subsequent 
BCS calculations. This provides the  triaxial qp-basis in the
present model.

The intrinsic states generated from the deformed Nilsson 
calculations don't conserve rotational symmetry. To  
restore this symmetry, three-dimensional angular-momentum projection technique 
is applied. From each intrinsic state, a band is generated through 
projection technique as discussed in Refs. \cite{HS79,HS80}. The interaction 
between different bands with a given spin is taken into account by diagonalising
the shell model Hamiltonian in the projected basis. 
The Hamiltonian used in the present work is given by 
\begin{equation}
\hat{H} = \hat{H_0} - \frac{1}{2} \chi \sum_\mu  \hat{Q}_\mu^\dagger 
\hat{Q}_\mu - G_M \hat{P}^\dagger \hat{P} 
- G_Q \sum_\mu \hat{P}_\mu^\dagger \hat{P}_\mu,
\label{Hamilt}
\end{equation}
with the corresponding mean-field (triaxial Nilsson) Hamiltonian 
\begin{equation}
\hat H_N = \hat H_0 - {2 \over 3}\hbar\omega\left\{\beta~\cos\gamma~\hat Q_0
+\beta~\sin\gamma~{{\hat Q_{+2}+\hat Q_{-2}}\over\sqrt{2}}\right\} ,
\label{nilsson}
\end{equation}
%where $\hat{H_0}$ is the spherical single-particle shell model Hamiltonian.
%The second, third  and fourth terms in Eq.~(\ref{Hamilt}) represent quadrupole-quadrupole, 
%monopole-pairing, and quadrupole-pairing interactions, respectively.
In the above equations, $\hat H_0$ is the spherical single-particle
Nilsson Hamiltonian \cite{Ni69}. The parameters of the Nilsson
potential are fitted to a broad
range of nuclear properties and is employed as a
mean-field potential.
The QQ-force strength, $\chi$, in Eq. (\ref{Hamilt})
is adjusted such that the physical
 quadrupole deformation $\beta$ is obtained as a result of the
 self-consistent mean-field HFB  condition. The relation is given by
\cite{KY95}:
\begin{equation}
\chi_{\tau\tau'} =
{{{2\over3}\epsilon\hbar\omega_\tau\hbar\omega_{\tau'}}\over
{\hbar\omega_n\left<\hat Q_0\right>_n+\hbar\omega_p\left<\hat
Q_0\right>_p}},\label{chi}
\end{equation}
where $\omega_\tau = \omega_0 a_\tau$, with $\hbar\omega_0=41.4678
A^{-{1\over 3}}$ MeV, and the isospin-dependence factor $a_\tau$ is
defined as
\begin{equation}
a_\tau = \left[ 1 \pm {{N-Z}\over A}\right]^{1\over 3},\nonumber
\end{equation}
with $+$ $(-)$ for $\tau =$ neutron (proton).
It is to be noted that the strengths in the TPSM are fixed as in the
original projected shell model approach with axial symmetry. 
 The monopole pairing strength $G_M$ (in MeV)
is of the standard formas$:$
%{\bf The pairing interaction strengths of the Hamiltonian are taken as follows
 %$:$ The $QQ$-force strength $\chi$ is adjusted such that the physical
 %quadrupole deformation $\epsilon$ is obtained as a result of the
% self-consistent mean-field HFB calculations.
%The monopole pairing
% strength GM is of the standard form as$:$}
\begin{equation}
G_{M} = \biggr( G_{1}\mp G_{2}\frac{N-Z}{A}\biggr) \frac{1}{A} \,(\rm{MeV}),
\label{gmpairing}
\end{equation}
where $- (+)$ sign applies to neutrons (protons).
%\begin{equation}
%G_M = {{G_1 - G_2{{N-Z}\over A}}\over A} ~{\rm for~neutrons,}~~~~
%G_M = {G_1 \over A} ~{\rm for~protons.}
%\label{pairing}
%\end{equation}
In the present calculations,
the pairing strength parameters, $G_1=20.12$ and $G_2=13.13$,
are adopted so that the observed odd-even mass differences
are reproduced with the model space of three active shells for
neutrons and protons. 
The active shells considered are $N=3,4,5$ for protons
  $N=4,5,6$  for neutrons.
The Nillson parameters $\kappa$ and $\mu$
for the major harmonic oscillator
shells have been adopted from Zhang et al. \cite{JZ87} for the present set of
calculations, and are listed in Table \ref{Tab:kmu}.
 The quadrupole pairing strength $G_Q$ is
assumed to be proportional to $G_M$ and the proportionality
constant being fixed as 0.16.  These parameters are similar to those used
in the earlier studies \cite{bh14a,bh15b}.
%===============================  table 0  ================================
\begin{table}
  \caption{ Set of Nilsson parameters used in the present calculation for odd-odd $^{194-200}$Tl-isotopes. }
\begin{tabular}{|c|cc|cc|c|}
  \hline       Major Shell (N)  & $\kappa_\pi$ &  $\mu_\pi$ & $\kappa_\nu$ &  $\mu_\nu$ \\ \hline
                3               &    0.090    &   0.300    &     -       &    -      \\ 
                4               &    0.065    &   0.570    &   0.070     &   0.390   \\ 
                5               &    0.060    &   0.540    &   0.062     &   0.430   \\ 
                6               &     -       &    -       &   0.062     &   0.340    \\ \hline
 \end{tabular}
\label{Tab:kmu}
\end{table}
%===============================  table 0  ================================
\begin{table}
  \caption{ The deformation parameters ($\beta, \gamma$)
    employed in the calculation for
odd-odd $^{194-200}$Tl-isotopes. The  deformation parameters 
have been taken
from Refs. \cite{sb17,pl13,HP12,ea08}.  }
\begin{tabular}{ccccccc}
\hline            & $^{194}$Tl &$^{196}$Tl & $^{198}$Tl & $^{200}$Tl \\
\hline
%$\epsilon$ & 0.142   & 0.150      & 0.192      & 0.225      & 0.172    & 0.209    \\
%       $\epsilon'$& 0.100   & 0.109      & 0.115      & 0.150      & 0.130    & 0.130 
%       $\gamma^0$& 35   & 36             &  30        & 33         & 31        & 34
       $\beta$      & 0.178   &  0.168      & 0.157 & 0.157    \\
       $\gamma$ & 30      &  29         &  31    & 34         
\\\hline
\end{tabular}
\label{Tab:Tcr}
\end{table}
%===============================================

%%===============  fig.2 =====================================================
%%\begin{figure}[!htb]
%%\includegraphics [totalheight=10cm,width=13.0cm]{194_196Tl_energy_apri5.pdf} \caption{(Color
%%online) Comparison of experimental and the TPSM calculated energies
%%for $^{194,196}$Tl isotopes. }
%%\label{figte1}
%%\end{figure}

%===============================================================================
%%===============  fig.3 =====================================================
%%\begin{figure}[!htb]
%%\includegraphics [totalheight=10cm,width=13.0cm]{198_200Tl_energy_apri5.pdf} \caption{(Color
%%online) Comparison of experimental and the TPSM calculated energies
%%for $^{198,200}$Tl isotopes. }
%%\label{figte2}
%%\end{figure}

%===============================================================================

In the present study, we have also evaluated the transition
probabilities using the TPSM wavefunctions.
The reduced electric quadrupole transition probability $B(E2)$ from an initial state 
$( \sigma_i , I_i) $ to a final state $(\sigma_f, I_f)$ is given by \cite {su94}
\begin{equation}
B(E2,I_i \rightarrow I_f) = {\frac {e^2} {2 I_i + 1}} 
| \bra \sigma_f , I_f || \hat Q_2 || \sigma_i , I_i\ket |^2 .
\label{BE22}
\end{equation}
In the calculations, we have used the effective charges of 1.6e for protons 
and 0.6e for neutrons. 
The reduced magnetic dipole transition probability
$B(M1)$ is computed through 
\begin{equation}
B(M1,I_i \rightarrow I_f) = {\frac {\mu_N^2} {2I_i + 1}} | \bra \sigma_f , I_f || \hat{\mathcal M}_1 ||
\sigma_i , I_i \ket | ^2 , 
\label{BM11}
\end{equation}
where the magnetic dipole operator is defined as  
\begin{equation}
\hat {\mathcal {M}}_{1}^\tau = g_l^\tau \hat j^\tau + (g_s^\tau - g_l^\tau) \hat s^\tau . 
\end{equation}
Here, $\tau$ is either $\nu$ or $\pi$, and $g_l$ and $g_s$ are the orbital and the spin gyromagnetic factors, 
respectively. 
In the calculations
we use for $g_l$ the free values and for $g_s$ the free values 
damped by 0.85 factor, i.e.,
\begin{eqnarray}
&&g_l^\pi = 1, ~~~ 
g_l^\nu = 0, ~~~   
g_s^\pi =  5.586 \times 0.85, \nonumber\\
&&g_s^\nu = -3.826 \times 0.85.
\end{eqnarray}
The reduced matrix element of an operator $\hat {\mathcal {O}}$ ($\hat {\mathcal {O}}$ is either
$\hat {Q}$ or $\hat {\mathcal {M}}$) is given by
\begin{eqnarray}
%\begin{split}
\bra \sigma_f , I_f || \hat {\mathcal {O}}_L || \sigma_i , I_i\ket   
  = \sum_{\kappa_i , \kappa_f} {f_{I_i \kappa_i}^{\sigma_i}} {f_{I_f \kappa_f}^{\sigma_f}}\nonumber \\
 \times \sum_{M_i , M_f , M} (-)^{I_f - M_f}  
\left( \begin{array}{ccc}
 I_f & L & I_i \\
-M_f & M & M_i 
\end{array} \right) \nonumber \\
  \times \bra \phi_{\kappa_f} | {\hat{P}^{I_f}}_{K_{\kappa_f} M_f} \hat {\mathcal {O}}_{LM}
\hat{P}^{I_i}_{K_{\kappa_i} M_i} | \phi_{\kappa_i} \ket .
%\end{split} 
\end{eqnarray}
%=================fig1=============================================
\begin{figure*}[htb]
 \centerline{\includegraphics[trim=0cm 0cm 0cm
0cm,width=1.0\textwidth,clip]{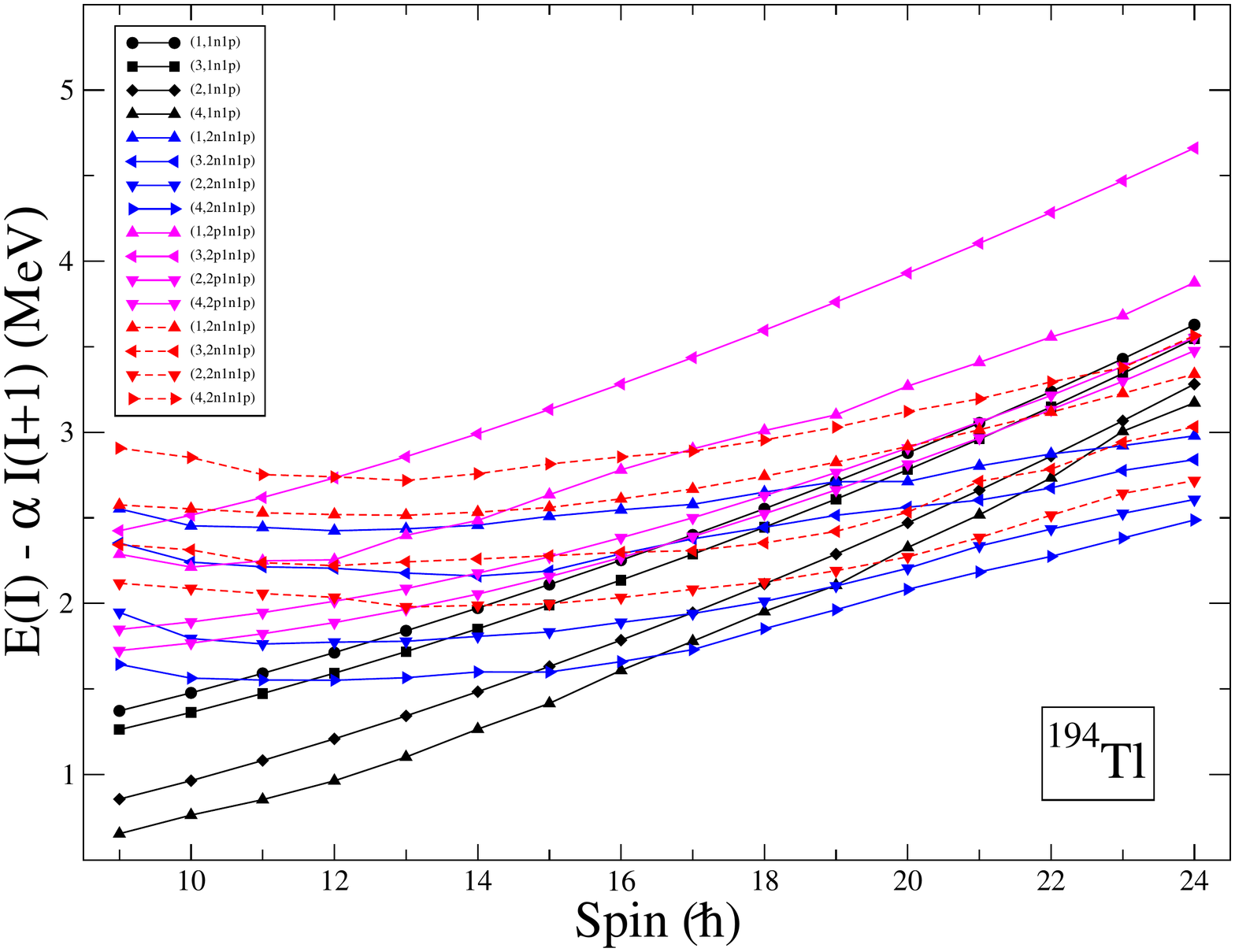}}
\caption{(Color online)  TPSM projected energies, before band mixing, of
  negative parity states for $^{194}$Tl. Bands are labelled by   $(K,\#)$
that designate the states with $K$ quantum number
and $\#$ the number of quasiparticles. For instance, (1,1n1p), (2,1n1p)
(3,1n1p), (4,1n1p) correspond to the $K=1,2,3,4$ one-neutron
coupled to one-proton
quasiparticle state. The value of $\alpha$, shown in y-axis, is defined as 
$\alpha = 32.32 A^{-5/3}$.} \label{figB1}
\end{figure*}
%=================================================

%=================fig1=============================================
\begin{figure*}[htb]
 \centerline{\includegraphics[trim=0cm 0cm 0cm
0cm,width=1.2\textwidth,clip]{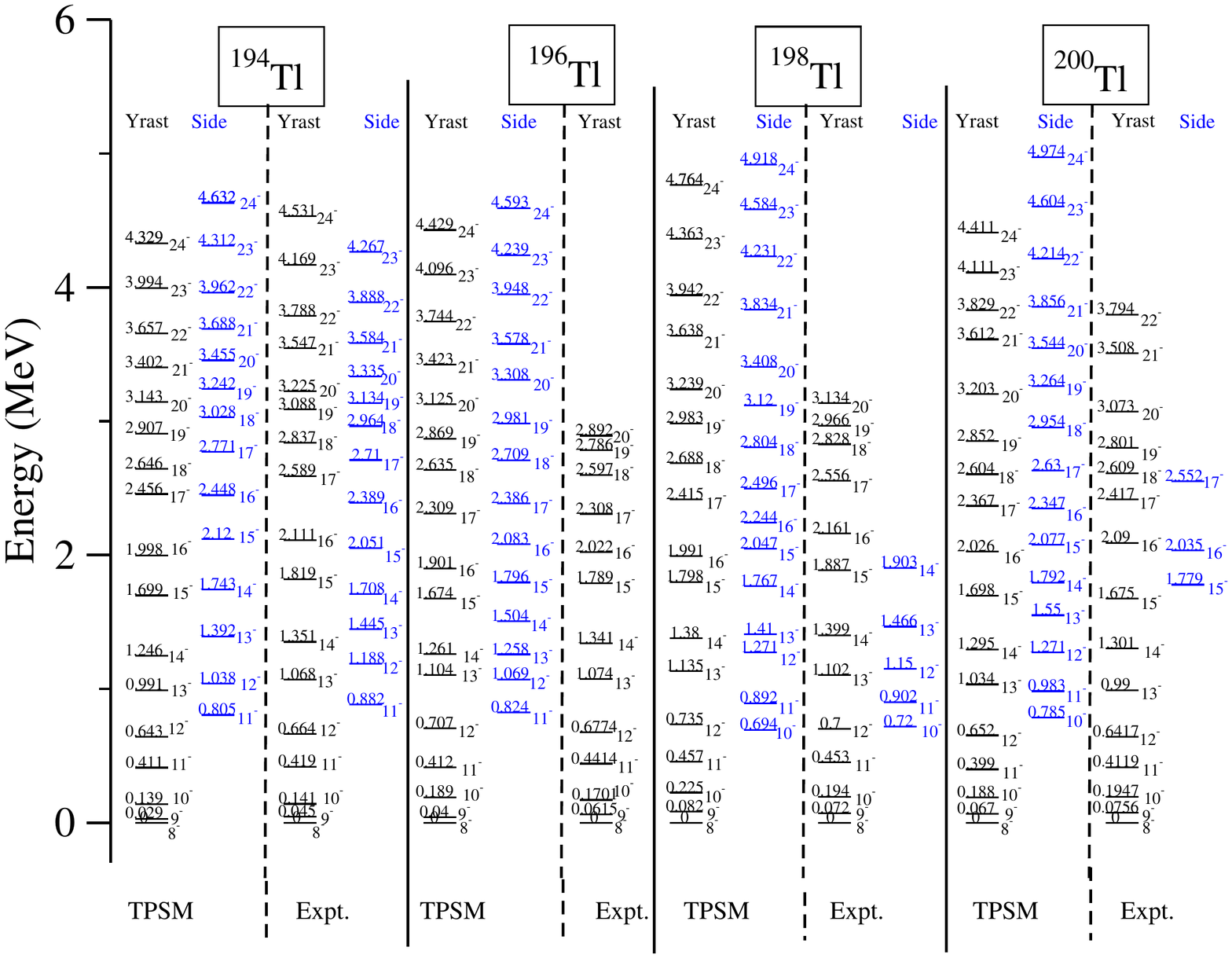}}
\caption{(Color online) Comparison of experimental and the TPSM calculated energies
for $^{194,196,198,200}$Tl isotopes. Experimental  results are from Refs.~\cite{sb17,pl13,HP12,ea08}. }
\label{figte1}
\end{figure*}
%==============================================================================

\section{Results and Discussions}
TPSM calculations have been performed for odd-odd $^{194-200}$Tl isotopes with
the deformation parameters listed in Table \ref{Tab:Tcr}. The axial and non-axial
deformation
values have been adopted from \cite{sb17,pl13,HP12,ea08}.
%and the non-axial deformations have been chosen (***).
Quasiparticle states are generated
with these deformation values by performing Nilsson plus BCS
calculations. The quasiparticle configurations close to the Fermi surface for
each system are then projected onto states with good angular momentum 
by performing three-dimensional angular momentum projection
\cite{RS80}. As an illustrative example, the lowest projected bands are
displayed in Fig. 1 for $^{194}$Tl. For other three isotopes, the projected
bands have a similar behaviour and are not discussed.
The ground state band for $^{194}$Tl in Fig.~\ref{figB1} is the projected band with K=4 from
one-proton coupled to one-neutron quasiparticle configuration having
quasiparticle energy of 1.9 MeV. This two quasiparticle configuration
is crossed at I=16 by 
two-neutron aligned state coupled to one-proton $\otimes$one-neutron
configuration having K=4 with quasiparticle energy of 3.5 MeV.

 The lowest projected bands, shown in Fig.~\ref{figB1}, and many more bands
close to the Fermi surface, which are about 40 for each nucleus, are
then used to diagonalize the shell model Hamiltonian of Eq.~\ref{Hamilt}. The
calculated lowest two bands, referred to as yrast and the side-bands,
are displayed in Fig.~\ref{figte1}  for the studied Tl isotopes. The available
experimental data is also shown in the  figure for 
comparison. For $^{194}$Tl, the data for both the bands is available
up to quite high spin and it is evident from Fig.~\ref{figte1} that TPSM
calculations reproduce the experimental results reasonably well. It is
noted that deviation for the highest observed I=24 state is only about 100 keV. 
The two observed bands have similar excitation energies and have been
proposed as a result of the chiral breaking machanism. This shall be 
discussed in the following when analysing the differences in  energies and transition
probabilities of the two bands in detail.

For $^{196}$Tl, the calculated yrast band is in good agreement with the
experimental data with the maximum deviation for I=20 state being
235 keV. The side band has not been observed for this
system and it is evident from the TPSM results that the predicted side
band deviates from the yrast band quite substantially as compared
to $^{194}$Tl. In the case of $^{198}$Tl, shown in Fig.~\ref{figte1}, yrast band is observed up to
I=20 and the side band only up to I=14. The TPSM calculations are
again noted to reproduce the data quite well up to the highest observed spin.
It is noted from the TPSM results that energy difference of the two
bands tend to decrease somewhat with increasing spin. The yrast band
in $^{200}$Tl is observed up to I=22 and only three states have been
populated for the side band. TPSM results reproduce the 
data fairly well.

\begin{figure}[!htb]
 \centerline{\includegraphics[trim=0cm 0cm 0cm
0cm,width=0.5\textwidth,clip]{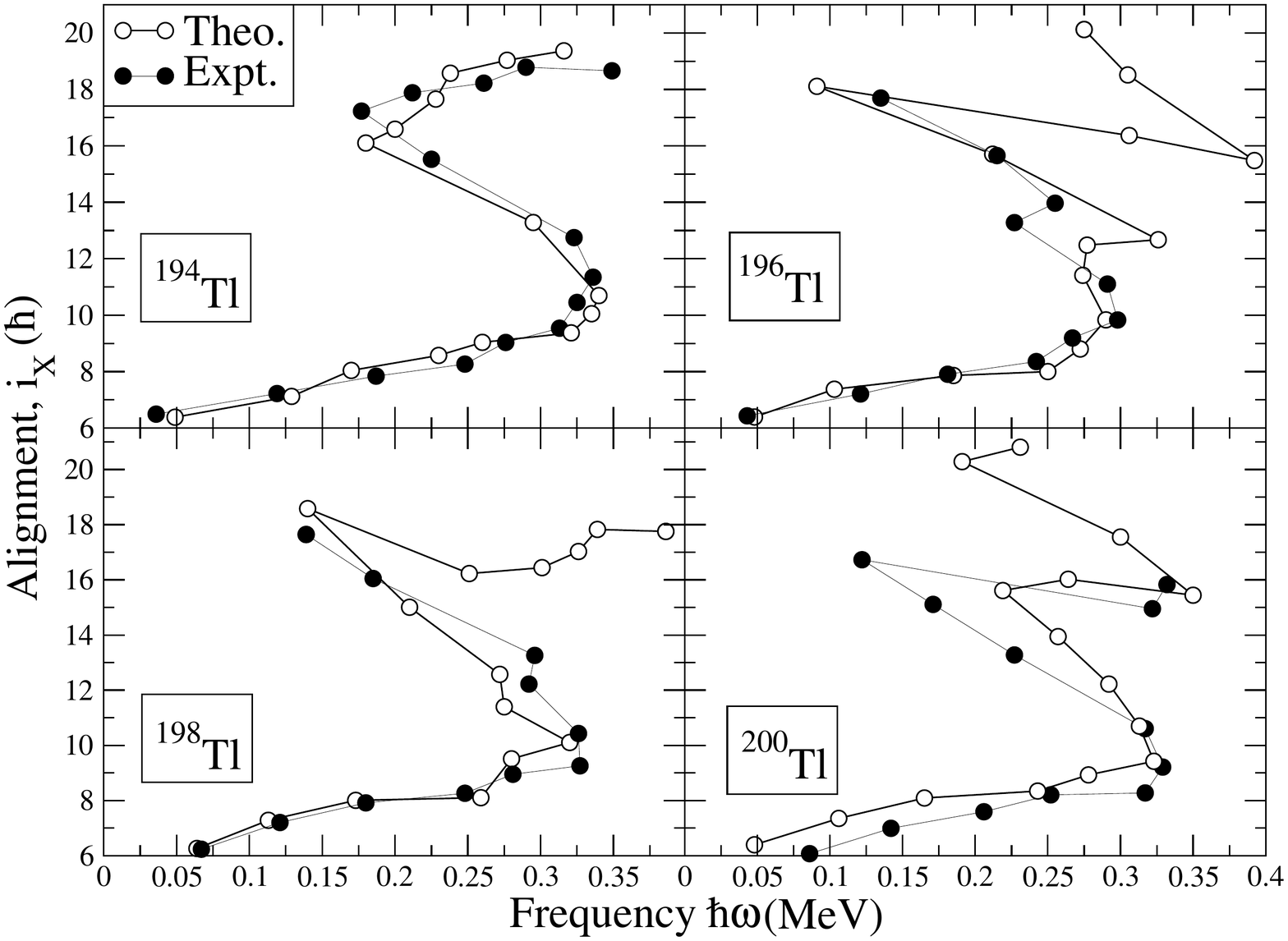}} \caption{(Color
online)  Comparison of experimental and the calculated alignment $I_x$ as a function of rotational frequency $\hbar\omega$ for the yrast bands of odd-odd $^{194-200}$Tl isotopes. The Harris reference parameters are taken as $J_0 = 8\hbar^2 MeV^{-1}$ and $J_1 = 40\hbar^4 MeV^{-3}$ \cite{HP12}.  }
\label{figi4}
\end{figure}

\begin{figure}[!htb]
 \centerline{\includegraphics[trim=0cm 0cm 0cm
0cm,width=0.5\textwidth,clip]{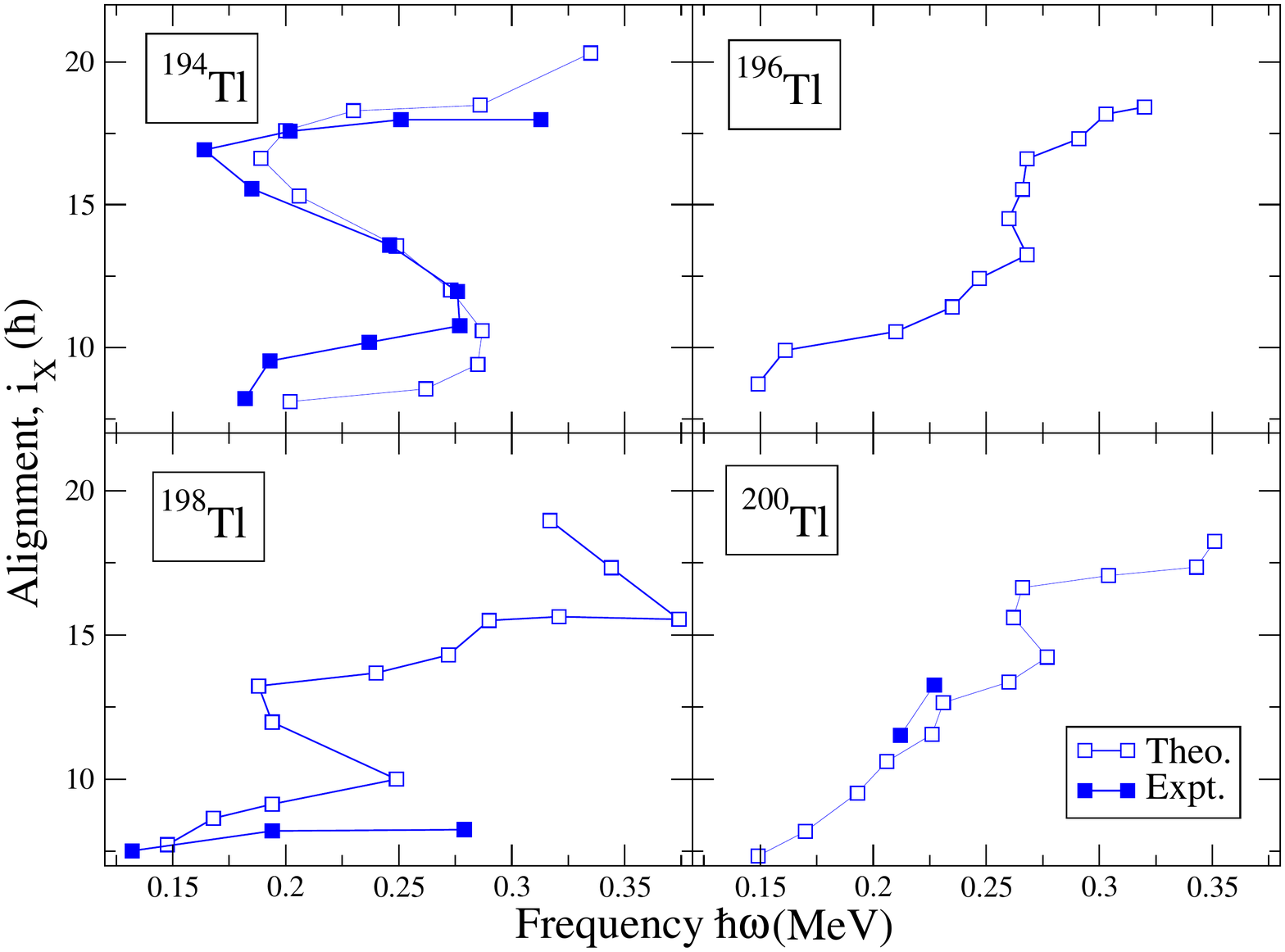}} \caption{(Color
online)  Comparison of experimental and the calculated alignment $I_x$ as a function of rotational frequency $\hbar\omega$ for the side bands of odd-odd $^{194-200}$Tl isotopes. The Harris reference parameters are taken as $J_0 = 8\hbar^2 MeV^{-1}$ and $J_1 = 40\hbar^4 MeV^{-3}$ \cite{HP12}.  }
\label{figi5}
\end{figure}

 To probe the crossing features of the observed band structures, the
aligned angular momentum $(i_x)$ for the yrast and the side bands are 
displayed in Figs.~\ref{figi4} and \ref{figi5}. It is apparent from the results
that TPSM calculations are in good agreement with the calculated values
from the experimental data. In
all the studied Tl-isotopes, the crossing is observed around 
0.3$ \hbar \omega$ and is due the alignment of two-neutrons.
This is evident from Fig. 1 with the configuration
having two-neutron coupled to one-proton $\otimes$ one-neutron crossing
the ground-state band.

\begin{figure}[!htb]
 \centerline{\includegraphics[trim=0cm 0cm 0cm
0cm,width=0.5\textwidth,clip]{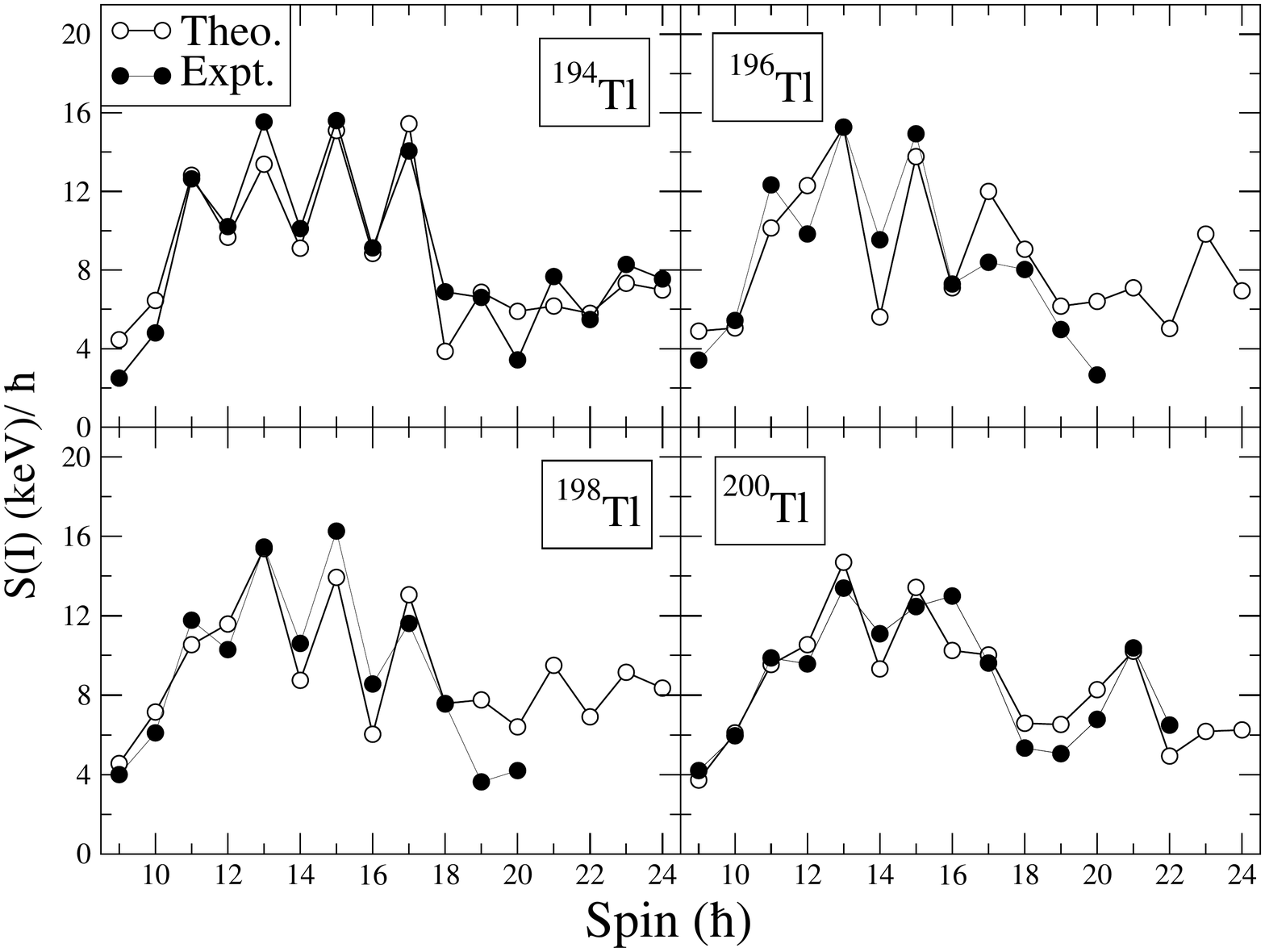}} \caption{(Color
online) Comparison of experimental and the calculated staggering $[S(I) = (E(I)-E(I-1))/2I]$ for the yrast bands of odd-odd $^{194-200}$Tl isotopes. }
\label{figs6}
\end{figure}

Further, deformed odd-odd nuclei in most of the regions of the
periodic table are known to depict
signature inversion in the low-spin regime. Considerable efforts have
been devoted to understand the nature of the signature inversion. It 
has been demonstrated that residual neutron-proton interaction and
triaxility can lead to signature inversion \cite{NT94}. To analyze whether
TPSM calculations are able to reproduce it, the energy staggering
defined by $S(I) = \frac{[E(I)-E(I-1)]}{(2I)}$ for the yrast bands are shown in
Fig.~\ref{figs6}. It is noted  from the figure that TPSM appoach is able to 
reproduce the observed signature inversion in $^{194}$Tl without
introducing any residual interaction. In other studies \cite{sb17,pl13,HP12,ea08}, additional
J-dependent neutron-proton interaction has been introduced to reproduce
the signature inversion. For other Tl-isotopes, it is seen
that TPSM results depicit a little delayed inversion point. However, the
magnitude of the signature splitting is reproduced fairly well for
the intermediate spin values and some deviation is noted at high spin
for $^{194}$Tl. The delayed signature inversion in $^{196-200}$Tl
isotopes needs further studies and in future we are planning to perform a
detailed analysis of the signature inversion in this mass region and
other regions using the TPSM approach.
%%===============  fig.4 =====================================================

%===============================================================================
%%===============  fig.5 =====================================================

%===============================================================================
%%===============  fig.6 =====================================================

%===============================================================================

%%===============  fig.7=====================================================
\begin{figure}[!htb]
 \centerline{\includegraphics[trim=0cm 0cm 0cm
0cm,width=0.5\textwidth,clip]{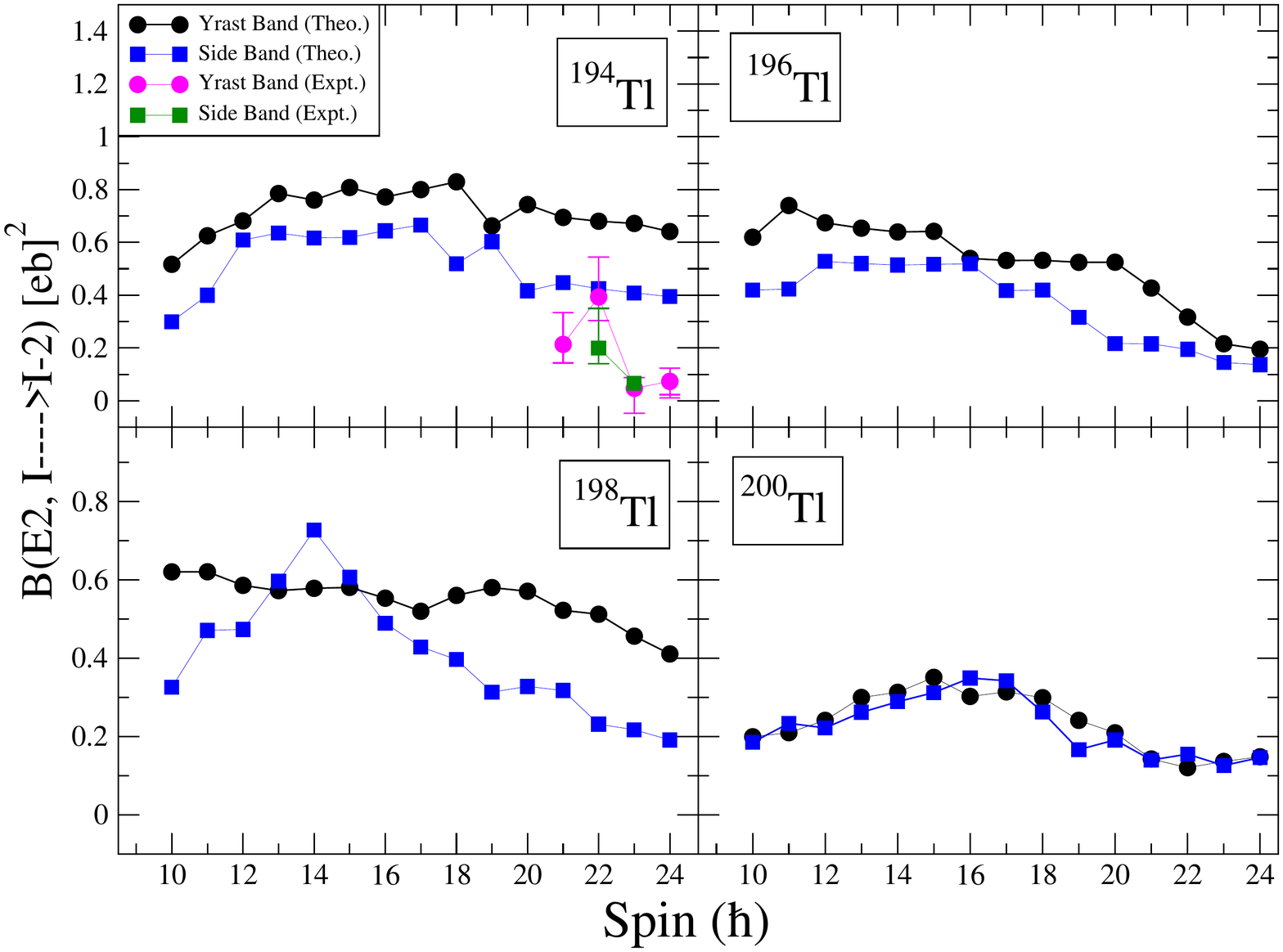}} \caption{(Color
online) TPSM Calculated values of  B(E2, I $\rightarrow$ I-2) ($e^2b^2$) for $^{194-200}$Tl isotopes. } \label{fige2}
\end{figure}
%==============================================================================

%%===============  fig.8 =====================================================
\begin{figure}[!htb]
 \centerline{\includegraphics[trim=0cm 0cm 0cm
0cm,width=0.5\textwidth,clip]{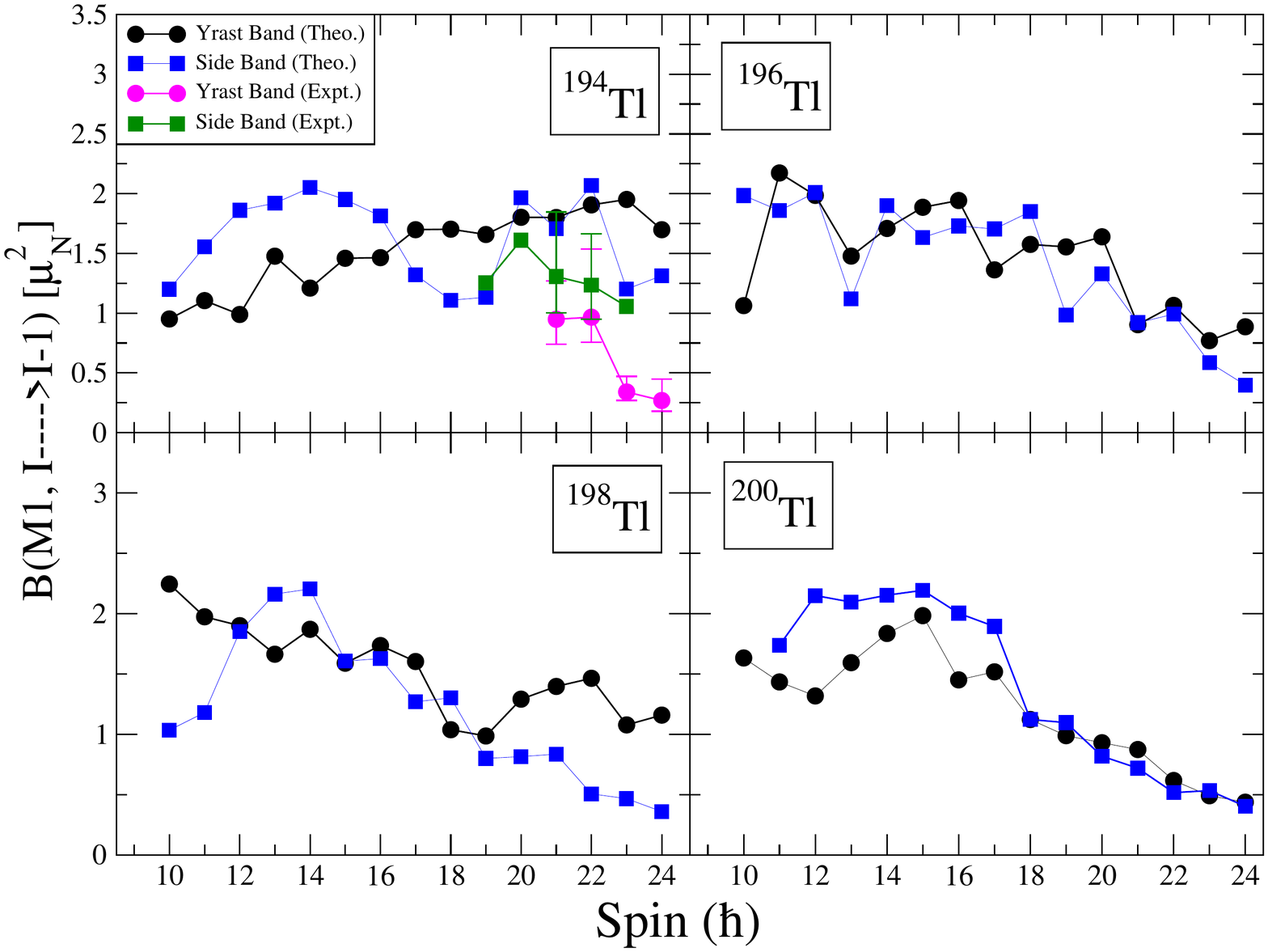}} \caption{(Color
online)  TPSM Calculated values of  B(M1, I $\rightarrow$ I-1) ($\mu^2_N$) for $^{194-200}$Tl isotopes.} \label{figM1}
\end{figure}
%==============================================================================

%%===============  fig.9 =====================================================
\begin{figure}[!htb]
 \centerline{\includegraphics[trim=0cm 0cm 0cm
0cm,width=0.5\textwidth,clip]{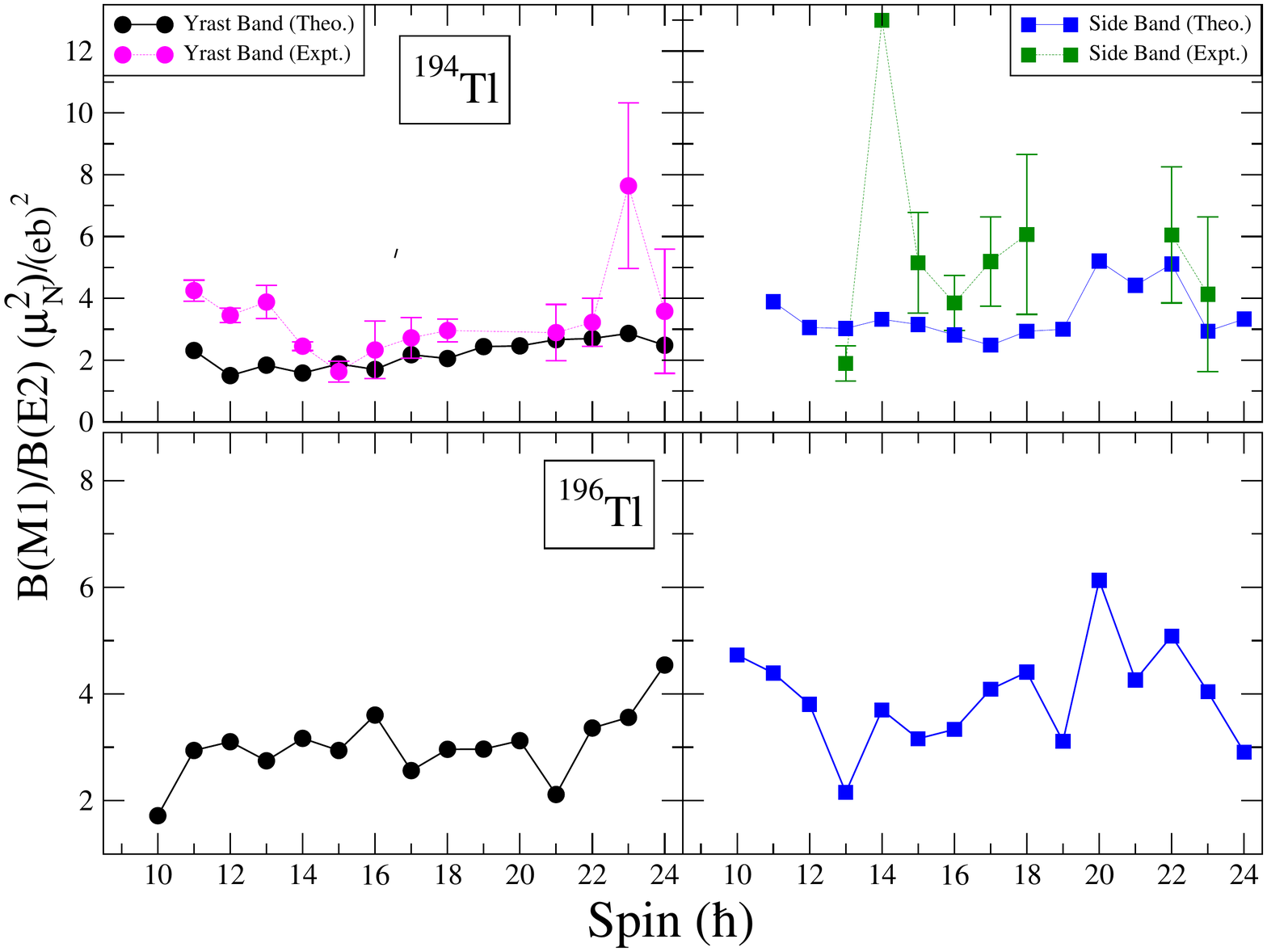}} \caption{(Color
online) Comparison of experimental and the calculated  B(M1)/B(E2) for $^{194,196}$Tl
isotopes. Experimental  results are from Refs.~\cite{pl13,ea08}.} \label{fig1m1e2}
\end{figure}
%==============================================================================
%%===============  fig.10=====================================================
\begin{figure}[!htb]
 \centerline{\includegraphics[trim=0cm 0cm 0cm
0cm,width=0.5\textwidth,clip]{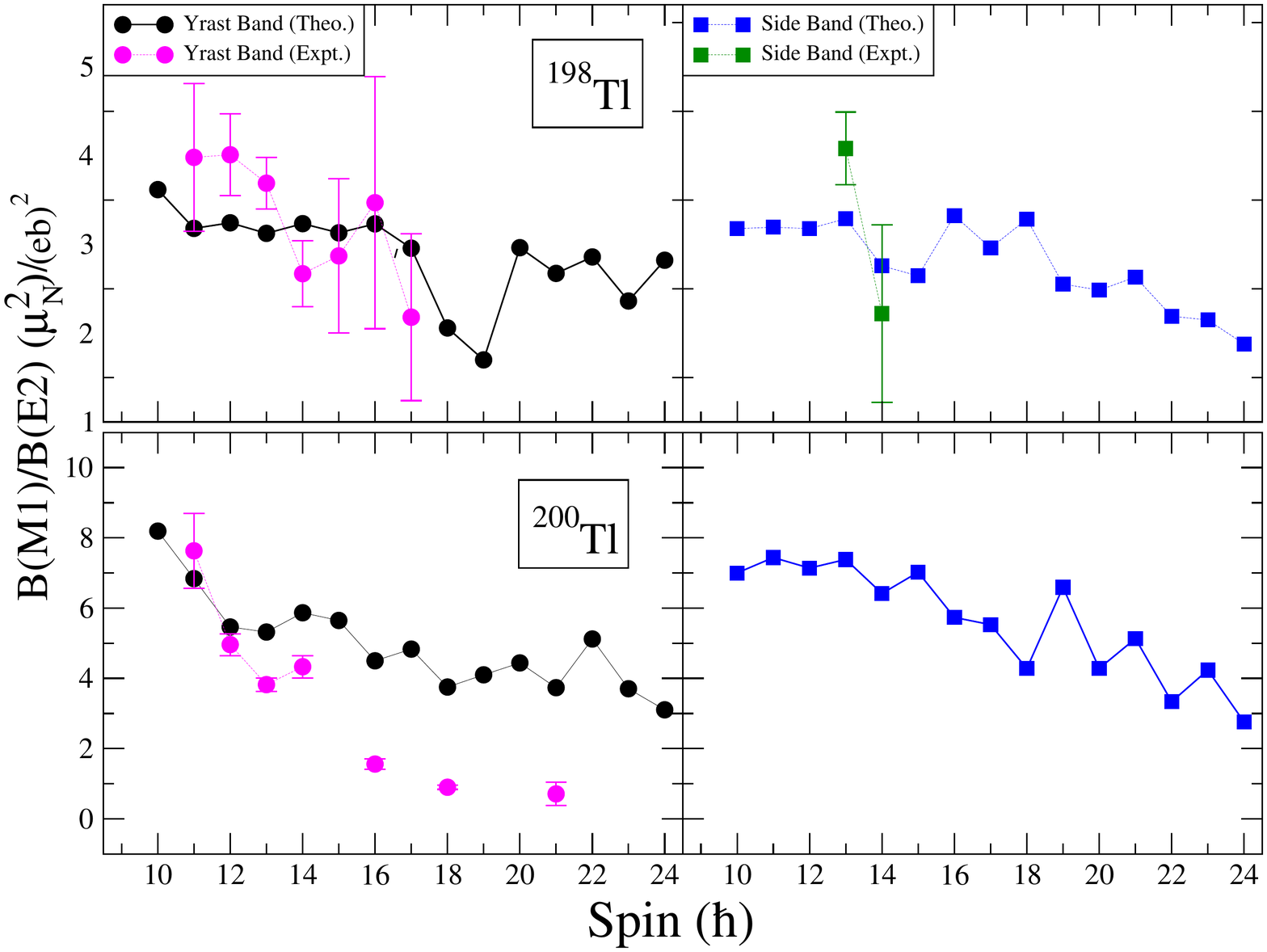}} \caption{(Color
online) Comparison of experimental and the calculated  B(M1)/B(E2) for $^{198,200}$Tl
isotopes. Experimental results are from Refs.~\cite{pl13,ea08}.} \label{fig2m1e2}
\end{figure}
%==============================================================================
%%===============  fig.9 =====================================================
\begin{figure}[!htb]
 \centerline{\includegraphics[trim=0cm 0cm 0cm
0cm,width=0.6\textwidth,clip]{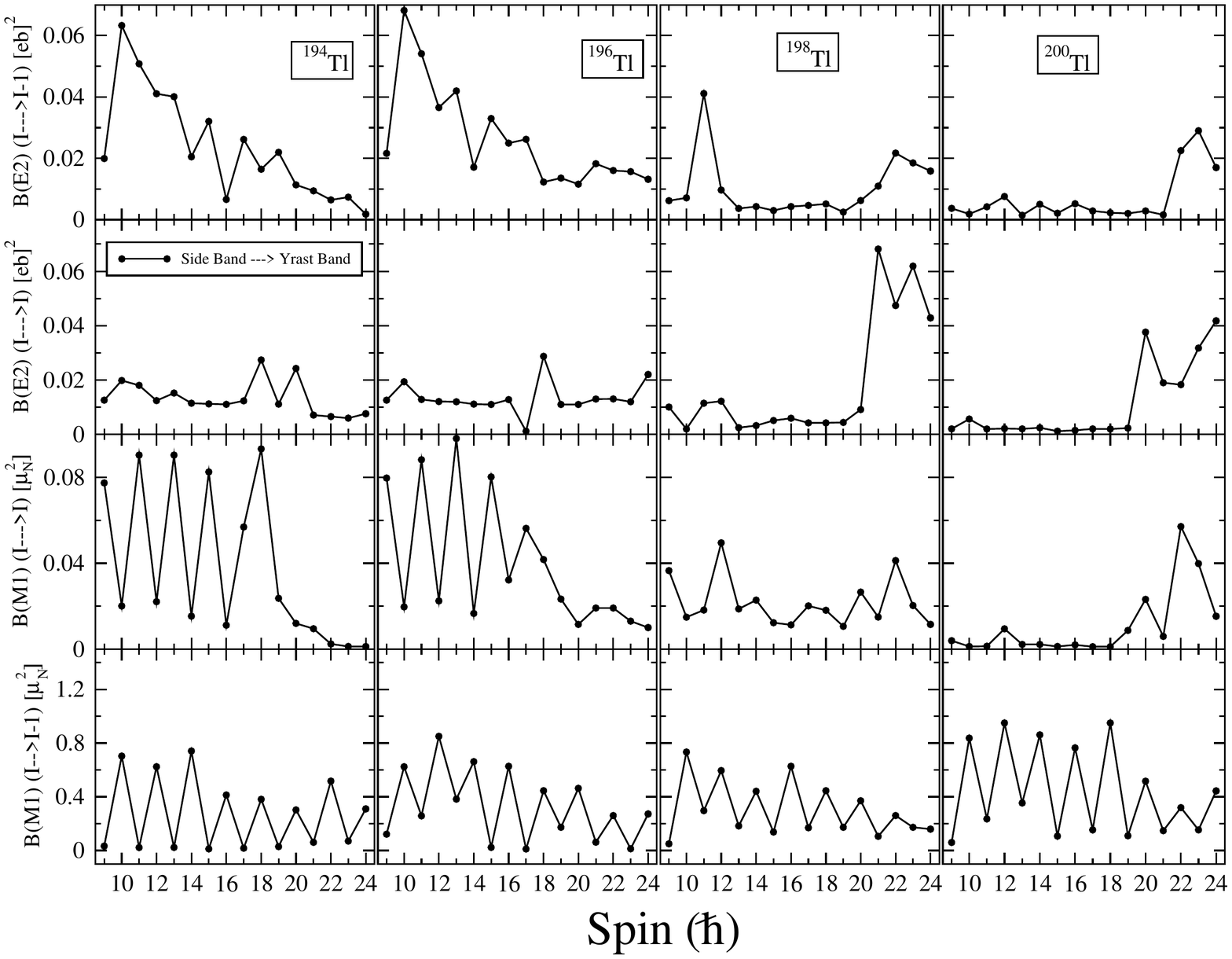}} \caption{(Color
   online) Calculated  $B(E2) (I \rightarrow I-1), B(E2) (I \rightarrow I), B(M1) (I \rightarrow I)$ and
   $B(M1) (I \rightarrow I-1)$  for $^{194-200}$Tl 
isotope from Side Band to Yrast Band. } \label{figMcbb}
\end{figure}
%==============================================================================

%%===============  fig.9 =====================================================
\begin{figure}[!htb]
 \centerline{\includegraphics[trim=0cm 0cm 0cm
0cm,width=0.6\textwidth,clip]{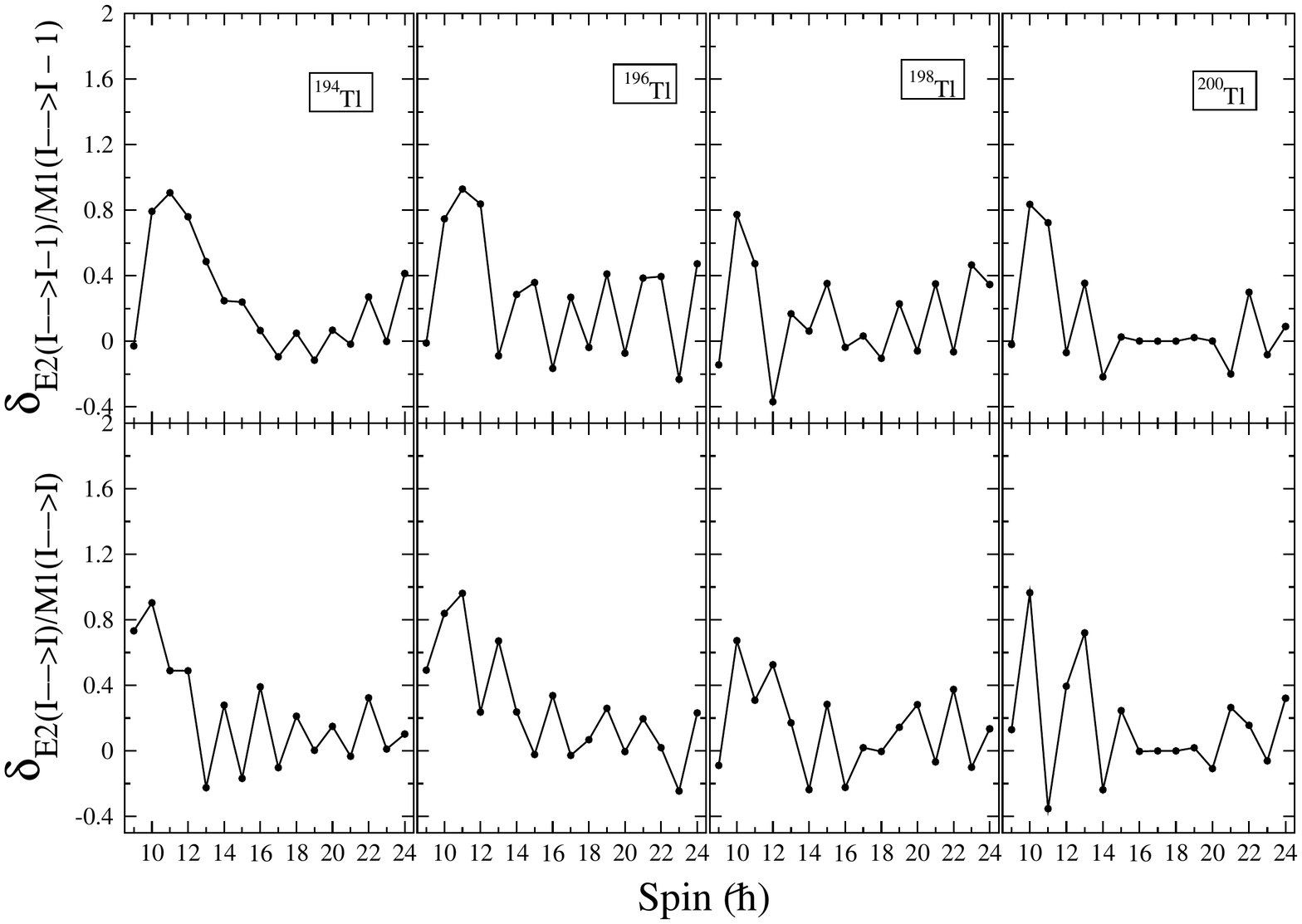}} \caption{(Color
   online) Mixing ratios for  $^{194-200}$Tl isotopes, calculated from
   the transitions given in Fig.~\ref{figMcbb} and using the expression
$~\delta_{E2/M1} = 0.835\,E_{\gamma}\,\frac{\bra I_f||E2||I_i\ket}{\bra I_f||M1||I_i\ket}$ \cite{FS82,mixing}.} \label{figMcb}
\end{figure}
%==============================================================================

%%===============  fig.11 =====================================================
\begin{figure}[!htb]
 \centerline{\includegraphics[trim=0cm 0cm 0cm
0cm,width=0.6\textwidth,clip]{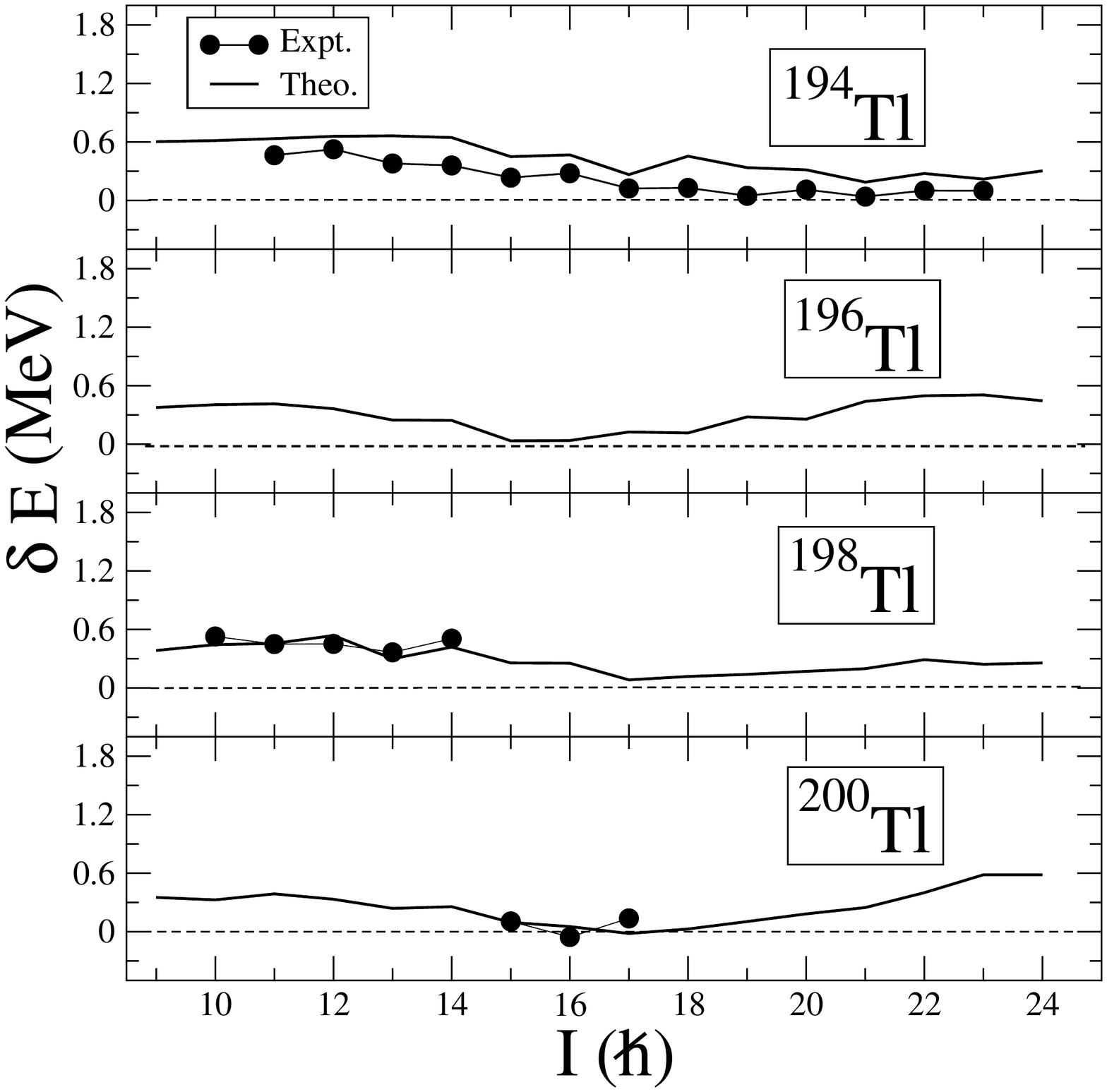}} \caption{(Color
online) 
Comparison of experimental and the calculated  Energy difference, $\delta (E)$,
between doublet bands as a function of spin. }
\label{figDeltaE}
\end{figure}
%====================================================
%%%%%%%%%%%%%%%%%%%%%%%%%%%%%%

%%%==================================
In the present work, we have also performed a detailed investigation of the transition
probabilities for the studied four Tl-isotopes. The effective charges used and
other relevant information have been presented in section II. The
calculated $B(E2)$ and $B(M1)$ transitions are displayed in
Figs.~\ref{fige2} and \ref{figM1}, respectively.  Unfortunately, except
for $^{194}$Tl, there is no
experimental data to compare with and we hope that 
data shall become available in the near future.

The data for the $B(M1)/B(E2)$ ratios is available for some bands and are
depicted in Figs.~\ref{fig1m1e2} and \ref{fig2m1e2} with the calculated TPSM values. For
$^{194}$Tl, ratios are available for both the yrast and the side band,
and it is seen  from the results that TPSM reproduces the
measured ratios quite well, except at I=23 for the yrast and at I=14
for the side band.  The experimental ratios are not available for
$^{196}$Tl and the TPSM ratios depicit similar behaviour as that for
$^{194}$Tl. The ratios available for $^{198}$Tl, shown in
Fig.~\ref{fig2m1e2}, are again noted to be in good agreement with the predicted
values.  The calculated TPSM ratios for $^{200}$Tl are seen to deviate
considerably from the experimental ratios at high spin. The
experimental numbers depicit a considerable drop with spin which is
not observed in the TPSM values. The reason for this discrepancy
could be due to fixed mean-field employed in the TPSM approach for all
the calculated spin values, which obviously is unrealistic. 

Though difficult to measure experimentally, we have
  calculated, using the TPSM wavefunctions, transitions between the doublet bands and
  the corresponding mixing ratios,
  and are displayed in Figs.~\ref{figMcbb} and  \ref{figMcb},
  respectively. It has been shown in a model study \cite{TS04} that these
  transitions  follow special selection rules for nuclei exibiting
  chiral geometry.  In Fig.~\ref{figMcbb}, it is evident that 
  inter-band $B(M1, I \rightarrow I)$ and $B(M1, I \rightarrow I-1)$ transitions have opposite phases for
  $^{194,196}$Tl-isotopes, which has been predicted in the model study
  as one of the characteristic features of the chiral symmetry. It is
  noted that mixing ratios are also measured \cite{mix} for chiral bands as $M1$
  and $E2$ transitions compete and are provided in Fig.~\ref{figMcb}
  for comparisons with future experimental measurements.
%%===============  fig.12 =====================================================
\begin{figure}[!htb]
 \centerline{\includegraphics[trim=0cm 0cm 0cm
0cm,width=0.5\textwidth,clip]{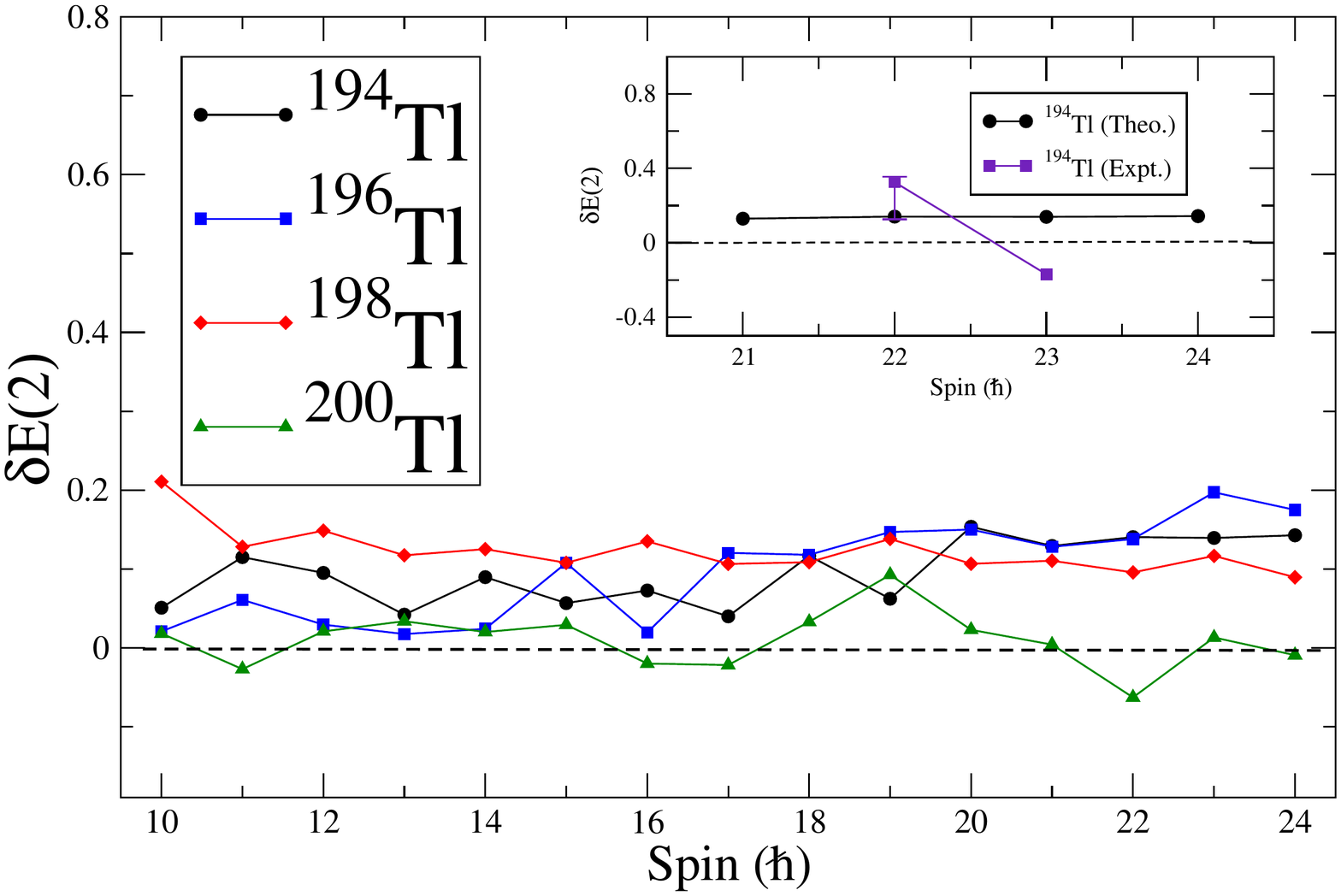}} \caption{(Color
online) 
 TPSM calculated  values of   $\delta$(E2) as a function of spin. }
\label{figDeltaE2}
\end{figure}
%====================================================

%%===============  fig.13 =====================================================
\begin{figure}[!htb]
 \centerline{\includegraphics[trim=0cm 0cm 0cm
0cm,width=0.5\textwidth,clip]{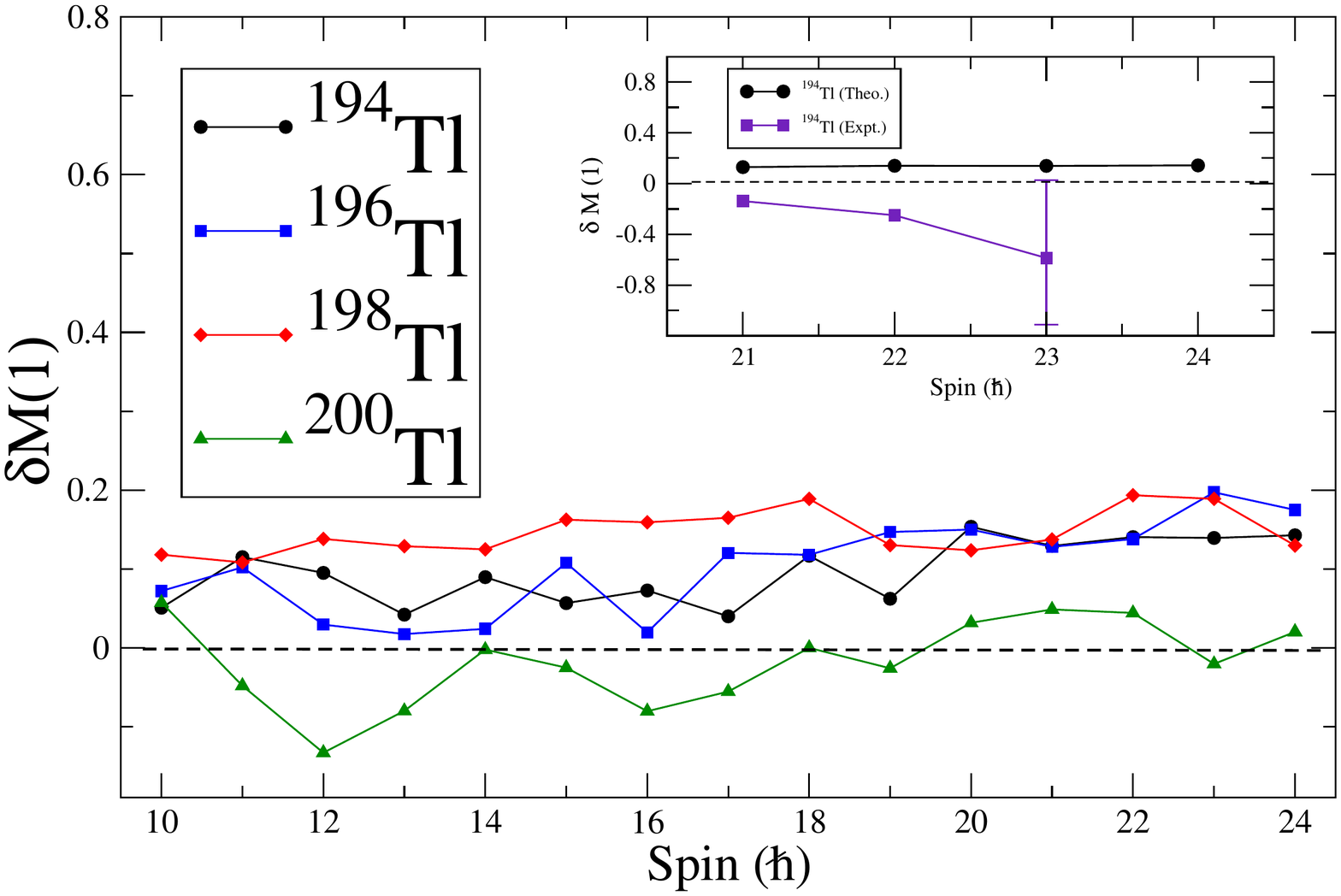}} \caption{(Color
online) 
 TPSM calculated  values of   $\delta$(M1) as a function of spin. }
\label{figDeltaM1}
\end{figure}
%====================================================

%%===============  fig.13 =====================================================
\begin{figure}[!htb]
 \centerline{\includegraphics[trim=0cm 0cm 0cm
0cm,width=0.5\textwidth,clip]{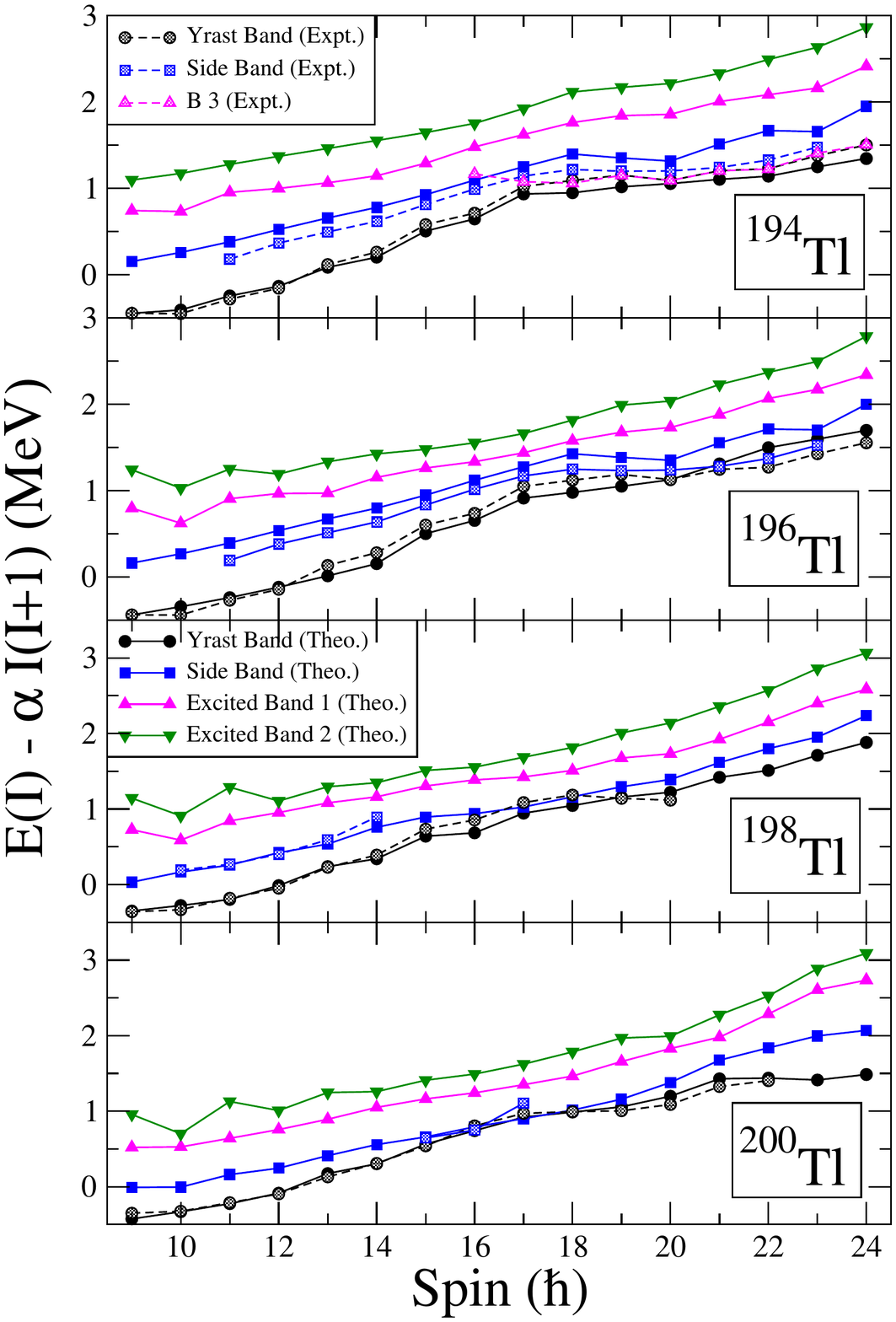}} \caption{(Color
online) Comparison of the measured energy levels of negative parity yrast and
excited bands for $^{194-200}$Tl isotopes.  The value of $\alpha$, shown in y-axis, is defined as 
$\alpha = 32.32 A^{-5/3}$.
 }
\label{Thexpt}
\end{figure}
%====================================================
%%===============  fig.13 =====================================================
\begin{figure}[!htb]
 \centerline{\includegraphics[trim=0cm 0cm 0cm
0cm,width=0.6\textwidth,clip]{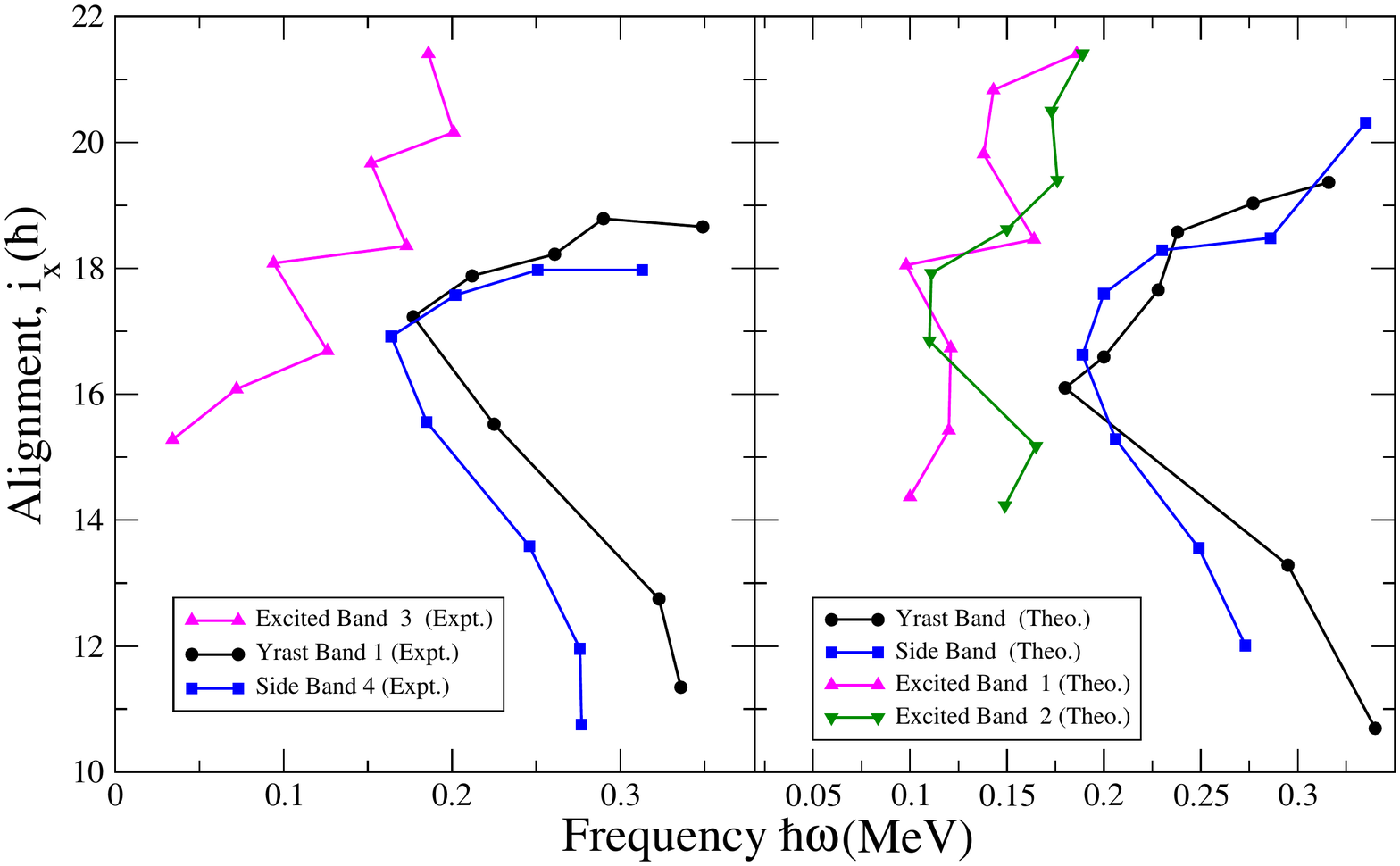}} \caption{(Color
online)  Comparison of experimental and the calculated alignment $I_x$ as a function of rotational frequency $\hbar\omega$ for the bands 3,  1 and 4 of $^{194}$Tl isotopes. The Harris reference parameters are taken as $J_0 = 8\hbar^2 MeV^{-1}$ and $J_1 = 40\hbar^4 MeV^{-3}$. Data is
taken from Ref. \cite{Masiteng}.
 }
\label{MOI194tl}
\end{figure}
%====================================================

%%===============  fig.9 =====================================================
\begin{figure}[!htb]
 \centerline{\includegraphics[trim=0cm 0cm 0cm
0cm,width=0.55\textwidth,clip]{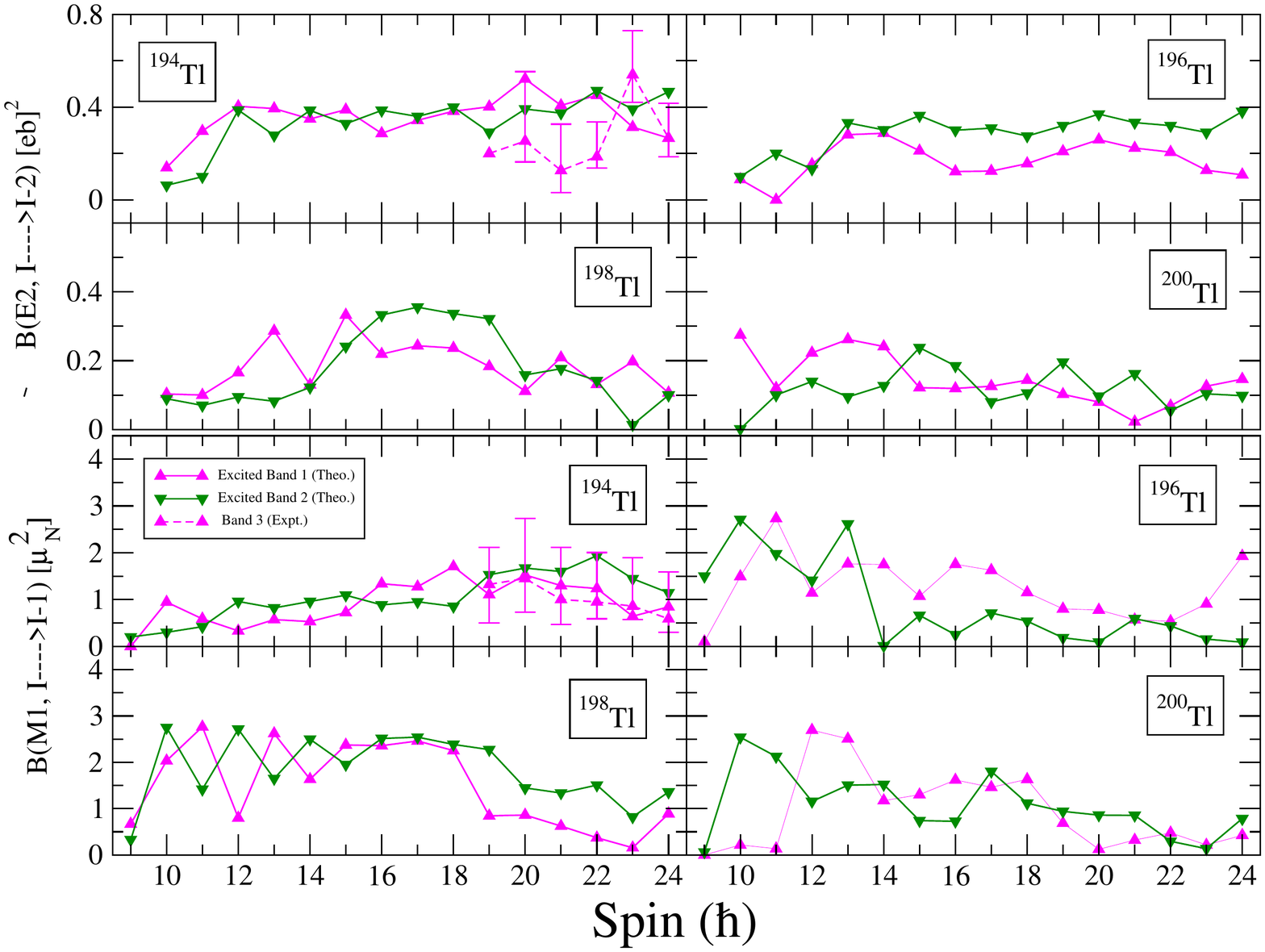}} \caption{(Color
online) Calculated  $B(E2) (I \rightarrow I-2)$ and $B(M1) (I \rightarrow I-1)$ for $^{194-200}$Tl isotopes. Experimental Data for $^{194}$Tl isotope is  from Ref.~\cite{Masiteng}.}
 \label{figMcb1}
\end{figure}
%==============================================================================

As mentioned in the introduction, it has been speculated \cite{sb17,pl13,HP12,ea08} that the doublet band structures observed in
Tl-isotopes could be  a result of the chiral symmetry breaking mechanism. Chiral
symmetry has been identified in many of the regions of the periodic
table and occurs in triaxially deformed systems with the angular momentum of
valence protons, valence neutrons, and that of the core aligned along
three orthogonal axis of the rotating nucleus. This leads to doublet
bands with identical spectroscopic properties. In order to investigate the
possibility of chiral symmetry breaking as the underpinning mechanism 
behind the emergence of doublet bands in Tl-isotopes,
we have evaluated the theoretical
and experimental differences of the energies and the transition probabilities
between the doublet bands in each of the three isotopes.
The difference in the transition probabilities is
defined as \cite{grodner}
\begin{equation}
\delta (\lambda\mu;I_{i}) =\frac{A-B}{A+B}  , \nonumber
\label{ctr1}
\end{equation}
where,
\begin{eqnarray}
&&A =\sqrt{(2I_{i}+1)B_{yrast}(\lambda\mu;I_{i}\rightarrow I)}  , \nonumber\\
&&B = \sqrt{(2I_{i}+1)B_{side}(\lambda\mu;I_{i}\rightarrow I)}\nonumber.
\label{ctr2}
\end{eqnarray}
It is expected that in the chiral symmetry breaking limit, the
differences of the calculated propreties for the doublet  bands should
tend to zero. The differences in the level energies [$\delta E$],
BE2 [$\delta E(2)$] and BM1 [$\delta M(1)$] 
are shown in Figs.~\ref{figDeltaE}, \ref{figDeltaE2} and \ref{figDeltaM1}, respectively.  For
$^{194}$Tl, both TPSM and the experimental $\delta(E)$ tend to zero with increasing
spin, in particular, after the band crossing at  I=18. It has
been postulated in the earlier work that four-quasiparticle band
structures, observed after the band crossing, could be the candidate
chiral bands \cite{pl13}. It is evident from Fig.~~\ref{figDeltaE} that for $^{194}$Tl, 
$\delta (E)$ does, indeed, tend to zero. For other studied Tl-isotopes,
the side bands have not been observed after the band
crossing. However, the TPSM calculated differences indicate that
$^{198}$Tl could possibly be another candidate of chiral symmetry 
breaking. In the case of $^{196}$Tl and $^{200}$Tl isotopes, the calculated 
$\delta (E)$, as a matter of fact, diverges after the bandcrossing and,
therefore, chiral symmetry can be ruled out for these two
isotopes. However, we would like to caution that these differences could
also be an artifact of the model calculations and more experimental and
theoretical studies are required to make firm conclusions.

The calculated differences in the transition probabilities, shown in
Figs.~\ref{figDeltaE2} and \ref{figDeltaM1},  
although small in all the cases, but tend to deviate for high spin states in
$^{194}$Tl, $^{196}$Tl and $^{198}$Tl isotopes. For $^{200}$Tl, the
deviations somewhat oscillate about the dashed zero line. In the chiral limit,
these deviations are expected to approach zero with spin. 
%===========================

In a more recent experimental work \cite{Masiteng}, another negative parity dipole band
has been reported along with the previously known two chiral partner bands. The
measured transition probabilities reported in this work provide some
details on the nature of the third observed band. This band has also
been assigned to be built on the  $\pi h_{9/2} \otimes \nu i^{-3}_{13/2} $
configuration,
similar to the other two observed bands. Many particle rotor model
(MPRM) calculations \cite{Masiteng} suggest that the band 3 could be
one partner of the second chiral pair bands in $^{194}$Tl.  The
results of the MPRM model predict that the yrare partner of the first
chiral pair and the both the bands of the second chiral pair have
almost similar excitation energies. This predicted near degeneracy and the fact that
these bands are weakly populated, being non-yrast, could be the reason that
the partner band of the second chiral pair has not been observed experimentally.

The energies of the lowest four bands using the TPSM approach for the odd-odd 
Tl-isotopes studied in this work are depicted in Fig.~\ref{Thexpt}. It is
evident from these results that
third and fourth bands are close on energy and could be due to the
second chiral pair as predicted in MPRM study \cite{Masiteng}. However, it is known that
in odd-odd nuclei density of states is higher and the near degeneracy of
the bands could just be as a results of it. In order to shed some
light on the nature of the third observed band in $^{194}$Tl, the
alignments of the four bands, calculated using the TPSM approach, are
compared with those deduced from the observed energies of the three
known bands in Fig.~\ref{MOI194tl}. The alignment for the third observed band
is in close agreement with the alignment of the third excited band
predicted in the TPSM approach.  In the experimental work, the third band
becomes lower in energy than that of yrare at higher spin.

 The calculated B(E2) and B(M1) values for the excited
  two bands in the studied Tl-isotopes are shown in
  Fig.\ref{figMcb1}. It is quite evident from the figure that transitions are very
  similar for the two bands and may correspond to the excited chiral
  pairs. For  $^{194}$Tl,   the measured values for few spin states
  are in good agreement with the calculated values.  
Clearly, further investigations are
required to understand the excited band structures in Tl-isotopes.

\section{Summary and Conclusions}
In the present work, the TPSM approach for odd-odd nuclei has been 
generalized to include two-neutron and two-proton quasiparticle states
coupled to the basic one-neutron $\otimes$ one-ptoton. This allows to
investigate odd-odd nuclei beyond the first band crossing, which 
is delayed as compared to even-even systems 
due to blocking of the orbitals by the unpaired particles.
%However, the data has now become available up to quite high-spin, beyond the band crossing.
In odd-odd $^{194-200}$Tl isotopes, high-spin states have
been observed beyond the first band crossing and it has been proposed that the
doublet bands observed in some of these isotopes may be a result of 
chiral symmetry breaking mechanism. 

The generalized TPSM approach has  been employed
to investigate odd-odd $^{194-200}$Tl
isotopes and it has been shown that results are in fair
agreement with the experimental data, wherever available. The energy
differences for $^{194,200}$Tl tend to drop with increasing spin and lends
support to the conjecture that this near degeneracy may be due to the
chiral symmetry breaking machanism. However, the results of the
transition probabilities, calculated using the TPSM approach, deviate
from those expected in the chiral symmetry limit. It is, therefore, important
to perform the lifetime measurements of the doublet bands in order to
ascertain the true structure of these bands.

%==================

We would like to add that in a parallel effort \cite{FQ17,FQ18,yk19}, TPSM approach has been
extended to include more than four-quasiparticles states and
also K- and azimuthal- plots have been proposed to probe the chiral
nature of the doublet bands. It would be quite interesting to extend
such analysis to the Tl-isotopes studied in the present work.
 
\section*{ACKNOWLEDGEMENTS}
Three of us (GHB, JAS and NR) would like to acknowledge DST for
Project No.$CRG/2019/004960$ (Govt. of India) for providing the financial
support to carry out the research work.  The authors would also like to acknowledge
Dr. E.A. Lawrie and Dr. S. Bhattacharya for providing
the measured data on the electromagnetic transitions for $^{194,200}$Tl isotopes. 
%\section*{Acknowledgments}
%Add the reference of 195Tl


\begin{thebibliography}{9}
\bibitem{BM75}  A. Bohr and B. R. Mottelson, {\it  Nuclear Structure}, Vol.
  II (Benjamin Inc., New York, 1975).
  \bibitem{bm8} D. J. Rowe and J. L. Wood, {\it Fundamentals of
      Nuclear Models: Foundational Models}, (World Scientific, 2010).
  \bibitem{sb17}  Soumik Bhattacharya, S. Bhattacharyya, S. Das Gupta, H. Pai,
    G. Mukherjee, R. Palit, F. R. Xu, Q. Wu, A. Shrivastava, Md. A. Asgar, R. Banik,
    T. Bhattacharjee, S. Chanda, A. Chatterjee, A. Goswami, V. Nanal,
       S. K. Pandit, S. Saha, J. Sethi, T. Roy, and S. Thakur,  Phys. Rev. {\bf C 95}  (2017) 014301.
     \bibitem{pl13}P. L. Masiteng, E. A. Lawrie, T. M. Ramashidzha, R. A. Bark,
       B. G. Carlsson, J. J. Lawrie,
       R. Lindsay, F. Komati, J. Kau, P. Maine, S. M. Maliage, I. Matamba,
       S. M. Mullins, S. H. T. Murray, K. P. Mutshena, A. A. Pasternak,
       I. Ragnarsson, D. G. Roux, J. F. Sharpey-Schafer, O. Shirinda,
       P. A. Vymers, Phys. Lett. {\bf  B 719}  (2013) 83.
     \bibitem{HP12}   H. Pai, G. Mukherjee, S. Bhattacharyya, M. R. Gohil,
       T. Bhattacharjee, and C. Bhattacharya,   R. Palit, S. Saha,
       J. Sethi, T. Trivedi, Shital Thakur, B. S. Naidu, S. K. Jadav, and R. Donthi,   A. Goswami, S. Chanda,
    Phys. Rev. {\bf C 85}  (2012) 064313.
  \bibitem{ea08} E. A. Lawrie, P. A. Vymers, J. J. Lawrie, Ch. Vieu,
    R. A. Bark, R. Lindsay, G. K. Mabala, S. M. Maliage, P. L. Masiteng,
    S. M. Mullins, S. H. T. Murray, I. Ragnarsson, T. M. Ramashidzha,
    C. Schuck, J. F. Sharpey-Schafer, and O. Shirinda, Phys. Rev. {\bf C 78}  (2008) 021305(R).

   
%\bibitem{nilson}  S. G. Nilsson, Dan. Mat. Fys. Medd. {\bf 29}, 16 (1955).  
\bibitem{SF01}  S. Frauendorf, Rev. Mod. Phys. {\bf 73}  (2001) 463.
\bibitem{TS04}  T. Koike, K. Starosta, and I. Hamamoto, 
               Phys. Rev. Lett. {\bf 17}  (2004) 172502.
%%%%%%%%%%%%%%%%added&&&&&&&&&&&&&&
\bibitem{CHI}  J. Meng and S. Q. Zhang,  J. Phys. {\bf G 37}
   (2010) 064025.
\bibitem{TS06} J. Meng and P. Zhao, Phys. Scr., {\bf 91} (2016)
  053008.
%%%%%%%%%%%%%%%%%%%%%%%%%%%%%%%%%%%%%%%%%%
    \bibitem{SF97}  S. Frauendorf and  J. Meng, Nucl. Phys. {\bf A 617} (1997)  131.
 
    \bibitem{CV04}  C. Vaman, D. B. Fossan, T. Koike, K. Starosta,
      I. Y. Lee, and A. O. Macchiavelli,  Phys. Rev. Lett. {\bf 92}  (2001) 032501.
\bibitem{TS03}  T. Koike, K. Starosta, C. J. Chiara, D. B. Fossan, 
                and D. R. LaFosse, Phys.  Rev. {\bf C 67}  (2003) 044319.               
  \bibitem{KS01}  K. Starosta, T. Koike, C. J. Chiara, D. B. Fossan,
                D. R. LaFosse, A. A. Hecht, C. W. Beausang, M. A. Caprio,
                J. R. Cooper, R. Kr\"{u}cken, J. R. Novak, N. V. Zamfir,
                K. E. Zyromski, D. J. Hartley, D. L. Balabanski,
                Jing-ye Zhang, S. Frauendorf, and V. I. Dimitrov, Phys. Rev. Lett. {\bf 86} (2001) 971.

  \bibitem{AA01}  A. A. Hecht, C. W. Beausang, K. E. Zyromski,
                D. L. Balabanski, C. J. Barton, M. A. Caprio, R. F. Casten,
                J. R. Cooper, D. J. Hartley, R. Kr\"{u}cken, D. Meyer,
                H. Newman, J. R. Novak, E. S. Paul, N. Pietralla, A. Wolf,
                N. V. Zamfir, Jing-Ye Zhang, and F. D\"{o}nau, Phys. Rev. {\bf C 63}  (2001) 051302.
  \bibitem{SU03}  S. Zhu, U. Garg, B. K. Nayak, S. S. Ghugre, N. S. Pattabiraman, D. B. Fossan,
                T. Koike, K. Starosta, C. Vaman, R. V. F. Janssens, R. S. Chakrawarthy,
                M. Whitehead, A. O. Macchiavelli, and S. Frauendorf,
                Phys. Rev. Lett. {\bf 91}  (2003) 132501.
%%%%%%%%%%%%%%%%added&&&&&&&&&&&&&&

\bibitem{SUU1} C. Liu, S. Y. Wang, R. A. Bark, S. Q. Zhang, J. Meng,
  B. Qi, P. Jones, S. M. Wyngaardt, J. Zhao, C. Xu, S.G. Zhou,
  S. Wang, D. P. Sun, L. Liu, Z. Q. Li, N. B. Zhang, H. Jia, X. Q. Li,
  H. Hua, Q. B. Chen, Z. G. Xiao, H. J. Li, L. H. Zhu, T. D. Bucher,
  T. Dinoko, J. Easton, K. Juhász, A. Kamblawe, E. Khaleel,
  N. Khumalo, E. A. Lawrie, J. J. Lawrie, S. N. T. Majola,
  S. M. Mullins, S. Murray, J. Ndayishimye, D. Negi, S. P. Noncolela,
  S. S. Ntshangase, B. M. Nyak\'{o}, J. N. Orce, P. Papka,
  J. F. Sharpey-Schafer, O. Shirinda, P. Sithole, M. A. Stankiewicz,
  and M. Wiedeking,  Phys. Rev. Lett. {\bf 116}  (2016) 112501.
\bibitem{SUU3} B.W. Xiong and Y.Y. Wang. At. Data Nucl. Data Tables, {\bf 125} (2019) 193.
\bibitem{SUU} S. Y. Wang, B. Qi, L. Liu, S. Q. Zhang, H. Hua, X. Q. Li,
  Y. Y. Chen, L. H. Zhu, J. Meng, S. M. Wyngaardt, P. Papka, T.T.Ibrahim,
  R. A. Bark, P. Datta, E. A. Lawrie, J .J. Lawrie, S. N. T. Majola,
  P. L. Masiteng, S. M. Mullins, J. G\'{a}l, G. Kalinka,
  J. Moln\'{a}r, B. M. Nyak\'{o},
  J. Tim\'{a}r, K. Juh\'{a}sz, and R. Schwengner, Phys. Lett. {\bf B 703}
  (2011) 40.

%%%%%%%%%%%%%%%%%%%%%%%%%%%%%%%%%%%%%
 \bibitem{SU02} T. Roy, G. Mukherjee, Md. A. Asgar, S. Bhattacharyya, S. Bhattacharya, C. Bhattacharya, S. Bhattacharya, T. K. Ghosh, K. Banerjee, S. Kundu, T. K. Rana, P. Roy, R. Pandey, J. Meena, A. Dhal, R. Palit, S. Saha, J. Sethi, S. Thakur, B. S. Naidu, S. V. Jadav, R. Dhonti, H. Pai, A. Goswam, Phys. Lett. {\bf B 782}  (2018) 768.
                

% \bibitem{SZ03}S. Zhu, et al., Phys. Rev. Lett. {\bf 91},  132501 (2003).
 \bibitem{SZ05} S. J. Zhu, J. H. Hamilton, A. V. Ramayya, P. M. Gore, J. O. Rasmussen, V. Dimitrov, S. Frauendorf,
 R. Q. Xu, J. K. Hwang, D. Fong, L. M. Yang, K. Li, Y.J. Chen, X.Q. Zhang, E. F. Jones, Y. X. Luo,
 I. Y. Lee, W. C. Ma, J. D. Cole, M. W. Drigert, M. Stoyer, G. M. Ter-Akopian,
 and A. V. Daniel, Eur. Phys. J. A {\bf 25}   (2005) 459.   
                
\bibitem{VD00}  V. I. Dimitrov, S. Frauendorf, and F. D\"{o}nau, 
                Phys. Rev. Lett. {\bf 84}  (2000) 5732.
\bibitem{PO04} 	P. Olbratowski, J. Dobazewski, J. Dudek, and W. Pl\"{o}ciennik, 
           	Phys. Rev. Lett. {\bf 93}  (2004) 052501.
 \bibitem{PRM} S. Q. Zhang, B. Qi, S. Y. Wang, and J. Meng, 
                 Phys. Rev. {\bf  C 75}  (2007) 044307.
 \bibitem{RPA1} S. Mukhopadhyay, D. Almehed, U. Garg, S. Frauendorf, T. Li, P. V. Madhusudhana Rao,
                 X. Wang, S. S. Ghugre, M. P. Carpenter, S. Gros, A. Hecht, R. V. F. Janssens,
                 F. G. Kondev, T. Lauritsen, D. Seweryniak, and S. Zhu, Phys. Rev Lett. {\bf 99} (2007)  172501.
  \bibitem{RPA2} D. Almehed, F. D\"{o}nau, and S. Frauendorf, Phys. Rev.  {\bf C 83}  (2011) 054308.
  
  \bibitem{JS99}  J. A. Sheikh and K. Hara,
    Phys. Rev. Lett. {\bf 82}  (1999) 3968.
  \bibitem{JG12} G. H. Bhat, J. A. Sheikh, and  R. Palit, Phys. Lett. {\bf B 707}  (2012) 250.
 % \bibitem{bh14} G. H. Bhat, J. A. Sheikh W. A.~Dar, S. Jehangir, R.~Palit, and P. A. Ganai, 
    %              Phys. Lett. B {\bf 738}, 218 (2014).
 
\bibitem{bh14a}G. H. Bhat, R. N. Ali, J. A. Sheikh, and  R. Palit,
               Nucl. Phys. A {\bf  922}   (2014) 150.
\bibitem{bh15b}W. A. Dar, J. A. Sheikh,  G. H. Bhat, R. Palit,  
               R. N. Ali,  and S.~Frauendorf, 
               Nucl. Phys. A {\bf  933} (2015) 123.
               
% \bibitem{JG11} J. A. Sheikh, G. H. Bhat, Y.-X. Liu, F.-Q. Chen, and Y. Sun,
 %              Phys. Rev. {\bf C 84}, 054314 (2011).              
\bibitem{Chanli15} C. L. Zhang, G. H. Bhat, W. Nazarewicz, J. A. Sheikh
               and Y. Shi, Phys. Rev. C {\bf 92} (2015) 034307.
\bibitem{Chanli16} G. H. Bhat, J. A. Sheikh, Y. Sun,  and R. Palit
  Nucl. Phys. {\bf A 947} (2016) 127.
  \bibitem{Js16}  J. A. Sheikh,  G. H. Bhat, W. A. Dar, S. Jehangir and
P. A. Ganai, Phys. Scr. {\bf 91}  (2016)  063015.
%\bibitem{fgfg1}G. H. Bhat, J. A. Sheikh and R. Palit Phys. Lett. {\bf  B 707}, 83 (2012).
\bibitem{RS80} P. Ring and P. Schuck, {\it The Nuclear Many-Body Problem} 
  (Springer, New York, 1980).
  
  \bibitem{HS79} K. Hara and S. Iwasaki, Nucl. Phys. A {\bf  332} (1979) 61.
  \bibitem{HS80} K. Hara and S. Iwasaki, Nucl. Phys. A {\bf  348}  (1980) 200.
   
\bibitem{Ni69}  S. G. Nilsson, C. F. Tsang, A. Sobiczewski, Z. Szymanski,
                S. Wycech, C. Gustafson, I. Lamm, P. Moller, B. Nilsson,
                Nucl. Phys. A {\bf  131}, 1 (1969) .
\bibitem{KY95} K. Hara and Y. Sun, Int. J. Mod. Phys. {\bf E 4} (1995)  637.
      \bibitem{JZ87}  J. -Ye Zhang, A. J. Larabee and L. L. Riedinger, J. Phys. {\bf G 13} (1987) 75. 
\bibitem{su94}  Y. Sun and J.L. Egido, Nucl. Phys. A {\bf  580}  (1994) 1.  

\bibitem{NT94} N. Tajima, Nucl. Phys. {\bf A 572}  (1994) 365.
\bibitem{mix} S. Zhu, U. Garg, B. K. Nayak, S. S. Ghugre,
  N. S. Pattabiraman, D. B. Fossan, T. Koike, K. Starosta, C. Vaman,
  R. V. F. Janssens, R. S. Chakrawarthy, M. Whitehead,
  A. O. Macchiavelli, and S. Frauendorf, Phys. Rev. Lett. {\bf 91}  (2003) 132501.
  \bibitem{grodner} E. Grodner, Acta Physica Polonica B {\bf 39}  (2008) 531.
 
\bibitem{FS82} F. D\"{o}nau and S. Frauendorf in 'Proceedings of the
  International Conference on High. Angular Momentum Properties of
  Nuclei', Oak Ridge (July 1982), Nucl. Sci.
\bibitem{mixing} A. Gsannatsempo, G. Maino, A. Nannini, and P. Sona,
  Phys. Rev. C {\bf 48} (1993) 6.
\bibitem{Masiteng} P. L. Masiteng, A. A. Pasternak, E. A. Lawrie, O. Shirinda, J. J. Lawrie, R. A. Bark, S. P. Bvumbi, N. Y. Kheswa, R. Lindsay, E. O. Lieder, R. M. Lieder, T. E. Madiba, S. M. Mullins, S. H. T. Murray,
   J. Ndayishimye, S. S. Ntshangase, P. Papka, and J. F. Sharpey-Schafer,
   Eur. Phys.  J. A {\bf 52} (2016) 28.
%%%==========new refs as per sugestions of Ref. 1====
\bibitem{FQ17} F. Q. Chen, Q. B. Chen, Y. A. Luo, J. Meng, and S. Q. Zhang, Phys. Rev. C {\bf 96} (2017) 051303.
\bibitem{FQ18} F. Q. Chen, J. Meng, and S. Q. Zhang, Phys. Lett. B {\bf  785} (2018) 211.
\bibitem{yk19} Y. K. Wang, F. Q. Chen, P. W. Zhao, S. Q. Zhang, and  J. Meng, Phys. Rev. C {\bf 99} (2019) 054303.
\end{thebibliography}
\end{document}